\documentclass{elsarticle}

\usepackage{lineno,hyperref}
\modulolinenumbers[5]

\usepackage{graphics} 
\usepackage{epsfig} 
\usepackage{amsmath} 
\usepackage{amssymb}  

\usepackage{algpseudocode}
\usepackage{algorithm}
\usepackage{epstopdf}

\usepackage{url}

\usepackage{verbatim}

\usepackage{amsthm}
\usepackage{pgfplots}
\pgfplotsset{
	/pgfplots/stairs legend/.style={
		legend image code/.code={\draw [#1] (0,-0.03) |- (0.0233,0) |- (0.0467,0.05)|-  (0.07,0.02) -- (0.07,-0.03);
		},
	},
}

\usepackage{booktabs}
\usepackage{color}
\usepackage{cancel}

\usepackage{enumitem}

\usepackage{tikz}               
\usepgflibrary{arrows}

\def\JB{\textcolor{red}}
\usetikzlibrary{arrows,calc,positioning,decorations.markings,patterns}

\newtheorem{theorem}{\bf Theorem}

\newtheorem{Algorithm}[theorem]{\bf Algorithm}

\journal{Annual Reviews in Control}
\date{ \copyright  2020 Elsevier Ltd.  All rights reserved.}
\begin{document}

\begin{frontmatter}

\title{Robust and optimal predictive control of the COVID-19 outbreak\tnoteref{mytitlenote}}
\tnotetext[mytitlenote]{This work was supported by Deutsche Forschungsgemeinschaft (DFG, German
	Research Foundation) under Grants
	GRK 2198/1 - 277536708, AL 316/12-2 - 279734922.
	The authors thank the International Max
	Planck Research School for Intelligent Systems (IMPRS-IS) for supporting Lukas Schwenkel, Anne Koch, Julian Berberich, and Patricia Pauli.}

\author{Johannes K\"{o}hler, Lukas Schwenkel, Anne Koch, Julian Berberich, Patricia Pauli, Frank Allg\"{o}wer
}
\address{Institute for Systems Theory and Automatic Control, University of Stuttgart \\ E-mail addresses: $\{$johannes.koehler, lukas.schwenkel, anne.koch, julian.berberich, patricia.pauli, frank.allgower$\}$@ist.uni-stuttgart.de }

\begin{abstract}
We investigate adaptive strategies to robustly and optimally control the COVID-19 pandemic via social distancing measures based on the example of Germany.
Our goal is to minimize the number of fatalities over the course of two years without inducing excessive social costs.
We consider a tailored model of the German COVID-19 outbreak with different parameter sets to design and validate our approach.
Our analysis reveals that an open-loop optimal control policy can significantly decrease the number of fatalities when compared to simpler policies under the assumption of exact model knowledge.
In a more realistic scenario with uncertain data and model mismatch, a feedback strategy that updates the policy weekly using model predictive control (MPC) leads to a reliable performance, even when applied to a validation model with deviant parameters.
On top of that, we propose a robust MPC-based feedback policy using interval arithmetic that adapts the social distancing measures cautiously and safely, thus leading to a minimum number of fatalities even if measurements are inaccurate and the infection rates cannot be precisely specified by social distancing.
Our theoretical findings support various recent studies by showing that 1) adaptive feedback strategies are required to reliably contain the COVID-19 outbreak, 2) well-designed policies can significantly reduce the number of fatalities compared to simpler ones while keeping the amount of social distancing measures on the same level, and 3) imposing stronger social distancing measures early on is more effective and cheaper in the long run than opening up too soon and restoring stricter measures at a later time.
\end{abstract}

\end{frontmatter}

\section{Introduction}
\label{sec:introduction}

Social distancing is an effective way to contain the spread of a contagious disease, particularly when little is known about the virus and no vaccines or other pharmaceutical interventions are available. Social distancing and isolation (together with other non-pharmaceutical measures such as hygiene and face masks) have a direct influence on the infection rates and hence on the spread of the virus \cite{Maharaj2012,Kissler2020,Maier2020}. While this combination has proven effective during the last weeks, e.g. in the German outbreak of COVID-19, strict social distancing is also very costly in terms of economical and psychological damage, which naturally leads to a multi-objective decision problem. 

There have been numerous approaches to model the COVID-19 outbreak and to predict future behavior for different distancing policies in simulation studies. The most commonly used modeling approaches are different extensions of the SIR (susceptible--infected--removed) model formulated either as system dynamics or as agent-based simulations (e.g.\ for Germany \cite{Dehning2020_change_points_ger,Barbarossa2020,German2020}).
In many such studies, different policies are simulated and compared with respect to the goals that both the health care system is not overwhelmed
such that every patient in need receives treatment and the mortality rate is kept low,
and also such that the majority of people can resume social interaction as soon as possible. However, in line with \cite{Alleman2020} and others, we advocate to go from mere model predictions to (model predictive) control, since control generally offers \emph{the} theory to  develop and apply optimal or robust decision making under uncertainties. 

While mathematical modeling and control of epidemics is a topic with rich history (see, e.g., the survey in \cite{Nowzari2016} and the references therein), there have also been numerous approaches to apply control theory to the COVID-19 spread. In \cite{Casella2020}, the author applies control theoretic principles and insights to a simple model of the outbreak to point out the difficulties of the system at hand: fast unstable dynamics with significant delays. 
In more recent literature, multiple works have addressed the problem of open-loop optimal control for the COVID-19 pandemic. In \cite{shorten2020covid,Tsay2020}, for example, the authors argue in favor of 'on and off'-policies of the social distancing measures, yielding a bang-bang like optimal control strategy. Such 'on and off'-policies, where the control input switches between two states, however, could pose great challenges, amongst others, for the society, but also for production lines, supply chains and the economy in general.

In this paper, we propose optimal open-loop and feedback control strategies to handle the German COVID-19 outbreak.
We employ the recently developed SIDARTHE model~\cite{Giordano2020} in order to design control policies which minimize the number of fatalities within a time horizon of two years, without using excessive social distancing measures.
We also address robustness of our policies w.r.t.\ model and measurement uncertainties via a (robust) model predictive control (MPC) feedback strategy.
Note that the following discussion and results are all based on the information and data available prior to the initial submission of this paper (May 2020).

Similar to the setup in this paper, the authors in \cite{Demasse2020} explore the best policy to implement while waiting for the availability of a vaccine. In their paper, they also distinguish between varying severity of symptoms ('mild' or 'severe') and seek a solution to the multi-objective optimization problem of minimizing fatalities and costs due to the implementation of the control strategy itself. Their main outcome of the open-loop input strategy is qualitatively similar to our results in Section~\ref{sec:OC_3}: Start with a loose strategy, soon increase all distancing measures such that the health care system capacities are never extorted and then relax the social distancing measures gradually and slowly. 
Another example for an open-loop optimal policy applied to the COVID-19 pandemic is presented in~\cite{kantner2020beyond} where the authors consider optimal control of the German outbreak using a slightly simpler model as the one chosen in the present paper (without distinguishing between detected and undetected individuals), which also includes an increased mortality rate if the ICU capacity is exceeded. Therein, the objective is not only to minimize the number of fatalities but also the number of susceptible individuals at the end of the time horizon, thus aiming for herd immunity. 
Our investigations in Section~\ref{sec:nom_oc}, however, indicate that herd immunity cannot be reached in a reasonable amount of time without overwhelming the hospital capacities. Therefore, our approach minimizes the number of fatalities after two years, with the underlying assumption that a vaccine will be available thereafter. 

However, an open-loop optimal policy cannot suffice to control the COVID-19 pandemic given all the uncertainties in the spreading of the virus and the disease progression, as we will see in the numerical results. We argue, similar to \cite{Alleman2020}, that an MPC-based feedback strategy is
the right tool to develop optimal and robust social distancing policies, especially in the presence of model inaccuracies. 
By using online measurements of the current outbreak, feedback inherently introduces robustness with respect to uncertainties and disturbances to the policy. We also robustify the feedback mechanism by introducing a robust MPC-based feedback strategy for uncertain state measurements which is crucial in a situation where only a limited amount of data is available and, for example, the number of the currently infected persons can only be estimated roughly by applying different studies. 

Our results are also in accordance with a very recently published joint strategy paper for Germany by authors from different German research institutions (Fraunhofer-Gesellschaft, Helmholtz Association, Leibniz Association and Max Planck Society) \cite{Hermann2020}. Firstly, they also state that reaching herd immunity without the availability of a vaccine would either exceed the health care capacities (with a resulting high mortality rate) or take several years (cf.\ our results in Section~\ref{sec:nom_oc}). Secondly, they state that the goal of wiping out the virus can only be a robust solution if this eradication would be a worldwide effort with very high social and economic costs (cf.\ our results in Section~\ref{sec:nom_oc_1_1}), which seems impossible to realize. Finally, they suggest an adaptive strategy for all policies influencing the infection rates with the goal to keep the spread of COVID-19 at bay while requiring the least possible restraints on the society and economy. With exactly this reasoning, we develop suitable control approaches in Section~\ref{sec:OC_3}, Section~\ref{subsec:roc_1}, and Section~\ref{subsec:roc_2} for such an 'adaptive' strategy. 

To summarize, our key contributions are the following:
\begin{itemize}
\item We extend the model in \cite{Giordano2020} by a mortality rate dependent on the state of the health care capacity and fit the parameters with data from Germany (Section~\ref{sec:modeling}).
\item We develop an optimization problem for finding the optimal input (in terms of setting infection rates) that minimizes the number of fatalities while keeping the costs occurring due to distancing measures low (Section~\ref{sec:nom_oc}). Moreover, we show that such an optimal input has significant advantages compared to simpler baseline policies.
\item We show that simply applying a precomputed (optimized) input is dangerous if the model is uncertain and explain why feedback is of utmost importance when dealing with such an unstable and uncertain system. Further, we demonstrate how such feedback can be incorporated via MPC, 
and we showcase the advantages of this control policy (Section~\ref{subsec:roc_1}).
\item We develop a robust MPC-based feedback strategy, which takes model inaccuracies, uncertain state measurements, and inexact inputs into account and can thus handle the COVID-19 outbreak cautiously and safely (Section~\ref{subsec:roc_2}).
\end{itemize}

Although based on a simple model fitted with limited data, we hope that these high-level insights inspire further investigations, possibly on more complex epidemiological models, and can ultimately help decision makers to improve and optimize their policies to mitigate the spread of epidemics while keeping the toll on the society and economy low.

%
\section{Modeling of the COVID-19 epidemic}\label{sec:modeling}
In this section, we describe the model of the COVID-19 epidemic that we use for our subsequent control approach.
Our model is adapted from the SIDARTHE model proposed in~\cite{Giordano2020} with the key differences that (i) we use more recent data to estimate new parameters to model the German COVID-19 outbreak (in~\cite{Giordano2020}, the Italian outbreak was considered) and (ii) we model the fact that the mortality rate increases if the number of critically ill patients exceeds the capacity of the German health care system.
In Section~\ref{subsec:model}, we describe the model of~\cite{Giordano2020} and explain its ingredients.
Thereafter, in Section~\ref{subsec:parameters}, we provide details on our parameter estimation algorithm which fits the model to the German COVID-19 outbreak.
Finally, we propose an extension of the model by increasing the mortality rate when the health care system is overwhelmed in Section~\ref{subsec:mortality_rate}.

\subsection{The SIDARTHE model}\label{subsec:model}
\begin{figure}[hbtp]
\begin{center}
%
%
%

\definecolor{mycolor1}{rgb}{0.00000,0.44700,0.74100}%
\tikzset{>=latex}
\resizebox{0.8\textwidth}{!}{\begin{tikzpicture}
\node[circle,draw,fill=white!90!blue,minimum size = 3.2cm] (S) at (0,0) {};
\node at (0,1) {\Large\textbf{\JB{S}}};
\node at (0,0) {\textbf{\JB{S}USCEPTIBLE}};
\node[circle,draw,fill=white!90!blue,minimum size = 3.2cm] (I) at (6.5,0)  {};
\node at ($(I)+(0,1)$) {\Large\textbf{\JB{I}}};
\node at ($(I)+(0,0.45)$) {\textbf{\JB{I}NFECTED}};
\node at ($(I)+(0,0)$) {\small \textbf{asymptomatic}};
\node at ($(I)+(0,-0.35)$) {\small \textbf{infected}};
\node at ($(I)+(0,-0.7)$) {\small \textbf{undetected}};
\node[circle,draw,fill=white!90!blue,minimum size = 3.2cm] (D) at (13,0)  {};
\node at ($(D)+(0,1)$) {\Large\textbf{\JB{D}}};
\node at ($(D)+(0,0.45)$) {\textbf{\JB{D}IAGNOSED}};
\node at ($(D)+(0,0)$) {\small \textbf{asymptomatic}};
\node at ($(D)+(0,-0.35)$) {\small \textbf{infected}};
\node at ($(D)+(0,-0.7)$) {\small \textbf{detected}};
\node[circle,draw,fill=white!90!blue,minimum size = 3.2cm] (A) at ($(I)+(0,-4)$)  {};
\node at ($(A)+(0,1)$) {\Large\textbf{\JB{A}}};
\node at ($(A)+(0,0.45)$) {\textbf{\JB{A}ILING}};
\node at ($(A)+(0,0)$) {\small \textbf{symptomatic}};
\node at ($(A)+(0,-0.35)$) {\small \textbf{infected}};
\node at ($(A)+(0,-0.7)$) {\small \textbf{undetected}};
\node[circle,draw,fill=white!90!blue,minimum size = 3.2cm] (R) at ($(D)+(0,-4)$)  {};
\node at ($(R)+(0,1)$) {\Large\textbf{\JB{R}}};
\node at ($(R)+(0,0.45)$) {\textbf{\JB{R}ECOGNIZED}};
\node at ($(R)+(0,0)$) {\small \textbf{symptomatic}};
\node at ($(R)+(0,-0.35)$) {\small \textbf{infected}};
\node at ($(R)+(0,-0.7)$) {\small \textbf{detected}};
\node[circle,draw,fill=white!90!blue,minimum size = 3.2cm] (T) at ($(R)+(0,-4)$)  {};
\node at ($(T)+(0,1)$) {\Large\textbf{\JB{T}}};
\node at ($(T)+(0,0.45)$) {\textbf{\JB{T}HREATENED}};
\node at ($(T)+(0,0)$) {\small \textbf{acutely}};
\node at ($(T)+(0,-0.35)$) {\small \textbf{symptomatic}};
\node at ($(T)+(0,-0.7)$) {\small \textbf{infected}};
\node at ($(T)+(0,-1.05)$) {\small \textbf{detected}};
\node[circle,draw,fill=white!90!blue,minimum size = 3.2cm] (H) at ($(S)+(0,-4)$)  {};
\node at ($(H)+(0,1)$) {\Large\textbf{\JB{H}}};
\node at ($(H)+(0,0.45)$) {\textbf{\JB{H}EALED}};
\node at ($(H)+(0,0)$) {\small \textbf{detected}};
\node at ($(H)+(0,-0.37)$) {\small \textbf{or}};
\node at ($(H)+(0,-0.7)$) {\small \textbf{undetected}};
\node[circle,draw,fill=white!90!blue,minimum size = 3.2cm] (E) at ($(A)+(1,-5.5)$)  {};
\node at ($(E)+(0,1)$) {\Large\textbf{\JB{E}}};
\node at ($(E)+(0,0.45)$) {\textbf{\JB{E}XTINCT}};
\node at ($(E)+(0,-0.2)$) {\small \textbf{detected}};

\draw (S) edge[line width=1mm, color = blue,->] (I);
\node at ($0.5*(I)+0.5*(S)+(0,0.3)$) {\color{blue}\textbf{contagion}};
\node at ($0.5*(I)+0.5*(S)+(0,-0.35)$) {\color{blue}$\boldsymbol{\alpha I{\small +}\beta D{\small +}\gamma A{\small +}\beta R}$};
\draw (I) edge[line width=1mm, color = blue!40!black,->] (D);
\node at ($0.5*(I)+0.5*(D)+(0,0.3)$) {\color{blue!40!black}\textbf{diagnosis}};
\node at ($0.5*(I)+0.5*(D)+(0,-0.35)$) {\color{blue!40!black}$\boldsymbol{\epsilon}$};
\draw (A) edge[line width=1mm, color = blue!40!black,->] (R);
\node at ($0.5*(A)+0.5*(R)+(0,0.3)$) {\color{blue!40!black}\textbf{diagnosis}};
\node at ($0.5*(A)+0.5*(R)+(0,-0.35)$) {\color{blue!40!black}$\boldsymbol{\theta}$};
\draw (I) edge[line width=1mm, color = red!60!yellow,->] (A);
\node at ($0.5*(I)+0.5*(A)+(1.1,0.15)$) {\color{red!60!yellow}\textbf{symptoms}};
\node at ($0.5*(I)+0.5*(A)+(-0.35,0.15)$) {\color{red!60!yellow}$\boldsymbol{\zeta}$};
\draw (D) edge[line width=1mm, color = red!60!yellow,->] (R);
\node at ($0.5*(D)+0.5*(R)+(1.1,0.15)$) {\color{red!60!yellow}\textbf{symptoms}};
\node at ($0.5*(D)+0.5*(R)+(-0.35,0.15)$) {\color{red!60!yellow}$\boldsymbol{\zeta}$};
\draw (R) edge[line width=1mm, color = red!70!green,->] (T);
\node at ($0.5*(R)+0.5*(T)+(0.85,0.18)$) {\color{red!70!green}\textbf{critical}};
\node at ($0.5*(R)+0.5*(T)+(-0.37,0.16)$) {\color{red!70!green}$\boldsymbol{\mu}$};
\draw (A) edge[line width=1mm, color = red!70!green,->] (T);
\node at ($0.5*(A)+0.5*(T)+(0.41,0.44)$) {\color{red!70!green}\textbf{critical}};
\node at ($0.5*(A)+0.5*(T)+(-0.62,-0.07)$) {\color{red!70!green}$\boldsymbol{\mu}$};
\draw (T) edge[line width=1mm, color = red,->] (E);
\node at ($0.5*(T)+0.5*(E)+(-0.25,0.42)$) {\color{red}\textbf{death}};
\node at ($0.5*(T)+0.5*(E)+(0.35,-0.34)$) {\color{red}$\boldsymbol{\tau(T)}$};
\draw (I) edge[line width=1mm, color = green!70!gray,->] (H);
\node at ($0.5*(I)+0.5*(H)+(0.3,-1)$) {\color{green!70!gray}\textbf{healing}};
\node at ($0.5*(I)+0.5*(H)+(-0.23,0.3)$) {\color{green!70!gray}$\boldsymbol{\lambda}$};
\draw (A) edge[line width=1mm, color = green!70!gray,->] (H);
\node at ($0.5*(A)+0.5*(H)+(0,0.3)$) {\color{green!70!gray}$\boldsymbol{\kappa}$};
\draw (T) edge[line width=1mm, color = green!70!gray,->,out=170,in=-30] (H);
\node at ($(H)+(3.3,-1.25)$) {\color{green!70!gray}$\boldsymbol{\sigma(T)}$};
\draw (R) edge[line width=1mm, color = green!70!gray,out=-35,in=80] ($(T)+(2.2,-0.5)$);
\draw ($(T)+(2.2,-0.5)$) edge[line width=1mm, color = green!70!gray,out=260,in=0] ($(E)+(0,-1.9)$);
\draw ($(E)+(0,-1.9)$) edge[line width=1mm, color = green!70!gray,->,out=180,in=-67] (H);
\node at ($(H)+(2.25,-2.3)$) {\color{green!70!gray}\textbf{healing}};
\node at ($(H)+(2.33,-3.9)$) {\color{green!70!gray}$\boldsymbol{\kappa}$};
\draw (D) edge[line width=1mm, color = green!70!gray,out=-34,in=93] ($(T)+(2.85,2.5)$);
\draw ($(T)+(2.85,2.5)$) edge[line width=1mm, color = green!70!gray,out=-87,in=87] ($(T)+(2.85,0)$);
\draw ($(T)+(2.85,0)$) edge[line width=1mm, color = green!70!gray,out=-93,in=3] ($(E)+(4,-2.4)$);
\draw ($(E)+(4,-2.4)$) edge[line width=1mm, color = green!70!gray,out=183,in=-3] ($(E)+(-1,-2.4)$);
\draw ($(E)+(-1,-2.4)$) edge[line width=1mm, color = green!70!gray,->,out=177,in=-100] (H);
\node at ($(H)+(0,-3)$) {\color{green!70!gray}$\boldsymbol{\lambda}$};
\end{tikzpicture}}
%
\end{center}
\caption{Scheme of the considered model for the COVID-19 epidemic, adapted from~\cite{Giordano2020}.}
\label{fig:model}
\end{figure}
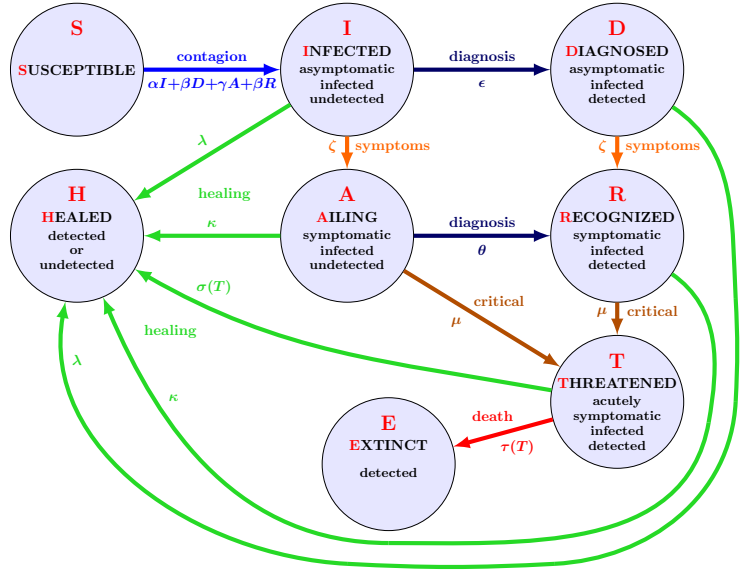

The considered model based on~\cite{Giordano2020} is shown in Figure~\ref{fig:model} and includes eight states:
\textbf{S - Susceptible}, \textbf{I - Infected} (asymptomatic, undetected), \textbf{D - Diagnosed} (asymptomatic, detected), \textbf{A - Ailing} (symptomatic, undetected), \textbf{R - Recognized} (symptomatic, detected), \textbf{T - Threatened} (symptomatic with life-threatening symptoms, detected), \textbf{H - Healed} (immune after prior infection, detected or undetected), \textbf{E - Extinct} (dead, detected). 
In accordance with Figure~\ref{fig:model}, the following differential equations describe the SIDARTHE model:
\begin{subequations}\label{eq:model}
\begin{align}\label{eq:model_S}
\dot{S}&=-S(\alpha I+\beta D+\gamma A+\beta R),\\\label{eq:model_I}
\dot{I}&=S(\alpha I+\beta D+\gamma A+\beta R)-(\epsilon+\zeta+\lambda)I,\\\label{eq:model_D}
\dot{D}&=\epsilon I-(\zeta+\lambda)D,\\\label{eq:model_A}
\dot{A}&=\zeta I-(\theta+\mu+\kappa)A,\\\label{eq:model_R}
\dot{R}&=\zeta D+\theta A-(\mu+\kappa)R,\\\label{eq:model_T}
\dot{T}&=\mu A+\mu R-(\sigma(T)+\tau(T))T,\\\label{eq:model_H}
\dot{H}&=\lambda I+\lambda D+\kappa A+\kappa R+\sigma(T) T,\\\label{eq:model_E}
\dot{E}&=\tau(T) T.
\end{align}
\end{subequations}
In the equations~\eqref{eq:model}, capital letters describe fractions of the whole population that are currently in the respective state.
Since the model represents the whole population, the states sum up to $1$, i.e., they must satisfy $S+I+D+A+R+T+H+E=1$ at all times.
Therefore, one equation in~\eqref{eq:model} is redundant and hence, e.g., the state $H$ can be expressed via the algebraic relation $H=1-(S+I+D+A+R+T+E)$ instead of Equation~\eqref{eq:model_H}, as it is common in the field of differential algebraic equations.
In most parts of this section, we omit time arguments for simplicity.
Further, Greek letters are the model parameters which are briefly summarized in the following:
\begin{itemize}
\item $\alpha,\beta,\gamma$ describe the infection rates for susceptible individuals, i.e., the rate at which susceptible individuals are infected by the states $I$, $D$ or $R$, and $A$, respectively, and hence join the state $I$.
\item $\epsilon,\theta$ describe the testing rate, i.e., at which rate (asymptomatic or symptomatic) infected individuals go from undetected to detected.
\item $\zeta$ describes the rate of asymptomatic (detected or undetected) infected individuals exhibiting symptoms, i.e., going from states $I$ or $D$ to $A$ or $R$, respectively.
\item $\mu$ is the rate at which infected individuals in $A$ or $R$ develop life-threatening symptoms, i.e., join the state $T$.
\item $\lambda,\kappa,\sigma(T)$ are recovery rates for individuals affected by COVID-19.
The recovery rate for threatened individuals $\sigma(T)$ depends on $T$, compare Section~\ref{subsec:mortality_rate}.
\item $\tau(T)$ is the mortality rate, i.e., the rate at which individuals with life-threatening symptoms decease, and it depends on $T$, compare Section~\ref{subsec:mortality_rate}.
\end{itemize}
Key features of the considered model for the COVID-19 pandemic compared to simpler ones (e.g., SIR models, compare~\cite{kermack1927contribution}) are that it distinguishes between detected and undetected cases, symptomatic and asymptomatic individuals, and it includes a separate state $T$ for patients with life-threatening symptoms (compare~\cite{Giordano2020} for a more detailed explanation of the key ingredients).
The present model, i.e., Equations~\eqref{eq:model} as well as Figure~\ref{fig:model}, is a mild modification of the model suggested in~\cite{Giordano2020}.
First, we reduce the number of parameters by including the following assumptions.
We assume that the rate for developing (severe) symptoms is the same for detected and undetected cases, since (to this day) no effective medication of COVID-19 is known.
More precisely, the transitions from $I$ to $A$ and $D$ to $R$ have the same dynamics with rate $\zeta$, and similarly for the respective recovery rates as well as for the transitions from $A$ to $T$ and $R$ to $T$.
Moreover, we assume that the rate $\beta$ at which susceptible individuals are infected is the same from states $D$ and $R$, since the state $D$ is neglected for the parameter identification step (compare Section~\ref{subsec:parameters}).
Finally, as a key difference to~\cite{Giordano2020}, we consider $T$-dependent rates $\tau(T)$ and $\sigma(T)$ for threatened patients, i.e., the mortality and recovery rates depend on the current number of threatened patients.
Essentially, $\tau(T)$ increases and $\sigma(T)$ decreases if $T$ exceeds the capacity of the German health care system (see Section~\ref{subsec:mortality_rate} for a detailed description of this model refinement).

\subsection{Parameters for the German outbreak}\label{subsec:parameters}
In this section, we adjust the model parameters and the initial condition given in \cite{Giordano2020} to the COVID-19 outbreak in Germany.
This is necessary, because the outbreaks in Germany and Italy evolved differently due to differences in the testing policy, the testing capacity, the health care system, the reaction of the governmental authorities, and the underlying counting method of confirmed cases.

In order to compute realistic parameters for Germany, we will use a pragmatic approach that enables us to easily include prior knowledge about relations between parameters.
The approach is a least squares optimization of the available data, where prior knowledge is incorporated via hard constraints in the optimization problem.
The available data is marked by a tilde and is given by:
\begin{itemize}
	\item the confirmed COVID-19 cases $\tilde C$, deaths $\tilde E$, and recoveries $\tilde H_c$ from \cite{Dong2020_jhu_dashboard}, \cite{data_jhu} for the days $t\in[0,49]$ from February 28, 2020 ($t=0$) to April 21, 2020 ($t=53$).
	We filtered this data set using the Matlab function \texttt{kaiser(7,3)} with window length $7$ and shape factor $3$ to reduce the effect of noise corruption and having less confirmed cases during weekends.
	Further, we have to divide the data set by the total German population $N_\mathrm{total}=8.3\cdot 10^{7}$ to ensure all values are normalized and are in the range $[0,1]$.
	\item the COVID-19 patients in ICU $\tilde T_2$ and how many of them died $\tilde E_2$ or recovered $\tilde H_2$ from \cite{divi2020report0404} for $t\in[24,53]$ from March 23 ($t=24$) to April 21 ($t=53$).\footnote{There was no reliable data available from before March 23; the data from March 23 to March 26 contains only information about $\tilde T_2$, not $\tilde E_2$ or $\tilde H_2$; from April 5 to April 7, there is a gap in the data due to a server migration of the DIVI Intensivregister \cite{divi2020report0404}.}
\end{itemize}
This data set, however, is rather small compared to the complexity of the model \eqref{eq:model} consisting of eight states and $13$ parameters. 
Therefore, we need to leverage additional prior knowledge in order to avoid over-fitting and ensure a realistic resulting parameter set and initial conditions.
Based on other studies and the interpretation of our model states and parameters, we enforce the following assumptions.
\begin{itemize}
	\item The detection rate of asymptomatic cases is negligible, as the current German policy is to test only people showing symptoms \cite{testPolicyGerZEIT}, i.e., $\epsilon=0$.
	\item At February 28, the initial date for our fit, there were $48$ confirmed cases, hence, we assume $R(0)=48/{N_\mathrm{total}}$, $T(0)=D(0)=H(0)=E(0)=0$, and $I(0)$, $A(0)$, $S(0)$ appear as decision variables with $S(0)=1-R(0)-I(0)-A(0)$.
	\item The test rate $\theta$ is approximately constant. Please note that this does not mean that the absolute number of tests is constant per day, as this value is rather proportional to $\theta A$ than to $\theta$.
	\item The infection rates $\alpha$ and $\gamma$ were influenced by the countermeasures that the German authorities installed to fight the  spread of the pandemic. According to \cite{Dehning2020_change_points_ger}, three main events changed the spreading rates: (1) March~9 --- canceling large events, (2) March~16 - closing schools and non-essential stores, and (3) March~23 --- contact ban (Kontaktsperre) that prohibits groups of more than two people and requires people to maintain a distance of at least $1.5\,\mathrm{m}$ in public. 
	Hence, there are four different policies $u_i$, $i=1,2,3,4$ monotonically increasing from no countermeasures $u_1=0$ to full lockdown $u_4=1$ resulting in $\alpha_i = \alpha_\text{max} + u_i(\alpha_\text{min}-\alpha_\text{max})$ and  $\gamma_i = \gamma_\text{max} + u_i(\gamma_\text{min}-\gamma_\text{max})$.
	This yields the following six decision variables $\alpha_\text{min}$, $\alpha_\text{max}$, $\gamma_\text{min}$, $\gamma_\text{max}$, $u_{2}$, and $u_3$.
	\item One of the main reasons why the COVID-19 pandemic is spreading so fast is that infectiousness peaks even before the onset of symptoms \cite{He2020}. 
	As asymptomatic individuals have no indication of their infection, they are on top of that also not as cautious as people with symptoms. 
	Therefore, we require $\alpha\geq \gamma$ when searching for realistic parameters. 
	Further, we want to ensure that people tested positive are significantly less contagious while in quarantine, such that we require $\gamma\geq 5\beta$.
	\item The percentage of \textit{confirmed} COVID-19 cases is estimated in the study \cite{Bommer2020} as $27.32\%$ in Germany.
	In our model, this value approaches the constant 
	$\phi= \frac{\zeta}{\lambda + \zeta} \frac{\theta + \mu}{\kappa+\theta+\mu}$, which is the proportion of people that develop symptoms ($I$ to $A$, $I$ to $D$ can be ignored as $\epsilon = 0$) and get detected ($A$ to $R$ or $T$); that is the percentage of confirmed accumulated cases in a steady state ($I=D=A=R=T=0$).
	To make sure our model coincides with the findings of \cite{Bommer2020}, we expect $\phi$ to be slightly above of the estimated $27.32\%$ as a steady state is not reached yet and the proportion of detected cases increases over time, i.e., we constrain $\phi \in [0.3, 0.45]$.
	\item The percentage of asymptomatic disease progressions was estimated at $43\%$ in a population screening study in Iceland \cite{Gudbjartsson2020}, at $43.2\%$ in a comprehensive testing of the whole municipality of Vo’, Italy \cite{Lavezzo2020} and at $17.9\%$ \cite{Mizumoto2020} in a study regarding the cruise ship Diamond Princess. 
	To ensure that our model has a comparable ratio, we add the constraint $\frac \lambda {\lambda + \zeta} \in [0.18, 0.43]$ to the optimization problem. 
	\item The (base) reproduction rate in the beginning of March was estimated as approximately $3$ \cite{Heiden2020}. Thus, for the parameters $\alpha_\text{max}$, $\gamma_\text{max}$ with no active countermeasures we require $R_0(\alpha_\text{max},\gamma_\text{max}) \in [2.5,3.5]$ where $R_0(\alpha, \gamma)$ is given by (see \cite{Giordano2020} for details)
	\begin{align}\label{eq:R_nought}
	R_0(\alpha, \gamma) = \frac{1}{\zeta+\lambda}\left(\alpha + \frac{1}{\theta+\mu+\kappa} \left( \gamma \zeta + \frac{\beta\theta\zeta}{\mu+\kappa}\right)  \right).
	\end{align}
	\item The median of the incubation time is $5$--$6$ days \cite{WHO_report_2020}, \cite{Linton2020}, \cite{Li2020}, which we identify as the half life period a person is in the state $I$, i.e. $\log(2)/(\lambda+\zeta)\in [5,6]$.
	Further, the median time from onset of symptoms until intensive care is $10$--$11$ days \cite{Yang2020}, \cite{Wang2020}. Hence, we constrain the half life period of a  ailing or recognized individuals to $\log(2)/(\kappa+\mu)\in [10,11]$.
\end{itemize}

\begin{figure}
	\centering
	\begin{tikzpicture}[node distance = 2cm, minimum width = 1cm, baseline=-1]
	\node [circle, draw] (AR) {\small $A\!+\!R$};
	\node [circle, draw, right of=AR, yshift=1cm] (T1) {$T_1$};
	\node [circle, draw, right of=AR, yshift=-1cm] (T2) {$T_2$};
	\node [circle, draw, right of=T1] (H) {$H$};
	\node [circle, draw, right of=T2] (E) {$E$};
	\draw [-latex] (AR) -- node[below]{$\mu_1$} (T1); 
	\draw [-latex] (AR) -- node[above]{$\mu_2$} (T2); 
	\draw [-latex] (T1) -- node[below]{$\sigma_1$} (H); 
	\draw [-latex] (T2) -- node[below, pos=0.85]{~~~~~$\sigma_2(T_2)$} (H); 
	\draw [-latex] (T1) -- node[above, pos=0.85]{~$\tau_1$} (E); 
	\draw [-latex] (T2) -- node[below]{$\tau_2(T_2)$} (E); 
	\end{tikzpicture}
	\quad$\approx$\quad
	\begin{tikzpicture}[node distance = 2cm, minimum width = 1cm, baseline=-1]
	\node [circle, draw] (AR) {\small $A\!+\!R$};
	\node [circle, draw, right of=AR] (T1T2) {$T$};
	\node [circle, draw, right of=T1T2, yshift=1cm] (H) {$H$};
	\node [circle, draw, right of=T1T2, yshift=-1cm] (E) {$E$};
	\draw [-latex] (AR) -- node[below]{$\mu$} (T1T2); 
	\draw [-latex] (T1T2) -- node[below]{~$\sigma(T)$} (H); 
	\draw [-latex] (T1T2) -- node[above]{~$\tau(T)$} (E); 
	\end{tikzpicture}
	\caption{The threatened state $T$ is split up in ICU cases $T_2$ and non-ICU cases $T_1$ which is later approximated using a lumped model. } \label{fig:T=T_1+T_2}
\end{figure}
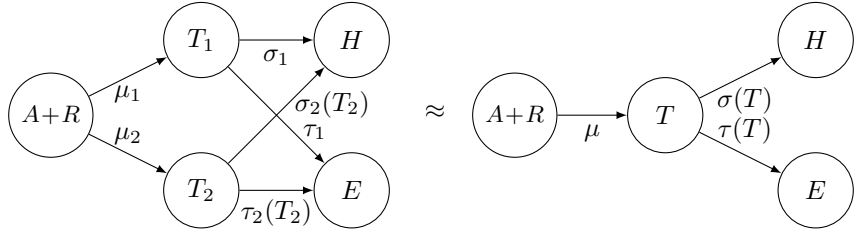

In the state $H$ of \eqref{eq:model} the confirmed recovered cases are not distinguished from the undetected ones, thus we define the number of confirmed recovered cases as $H_c$, with $H_c(0)=0$ and $\dot H_c=\lambda D+\kappa R + \sigma_1 T_1+\sigma_2T_2$ and further the number of confirmed accumulated cases as $C=D+R+T+H_c+E$ in order to match the data $\tilde H_c$ and $\tilde C$.
Considering the COVID-19 patients in intensive care $\tilde T_2$, a natural choice would be to identify them with threatened state $T$, however, all deaths in the model have been in $T$ before, but in reality only half to a third of the deaths happens in ICU \cite{data_jhu}, \cite{divi2020report0404}.
Hence, as the patients in ICU are only a part of $T$, we split $T$ into $T_1$ and $T_2$, where $T_2$ represents the number of people in intensive care and $T_1$ are all otherwise threatened COVID-19 cases.
We assume that there are no transitions from $T_1$ to $T_2$ and vice versa, such that $T=T_1+T_2$ can be modeled as
\begin{subequations}
	\begin{align}
	\dot T_1 &= \mu_1(A+R)-(\sigma_1+\tau_1)T_1\\ 
	\dot T_2 &= \mu_2(A+R)-(\sigma_1(T_2)+\tau_2(T_2)) 
	\end{align}
\end{subequations}
with $\mu_1+\mu_2 = \mu$. 
This more complex model with $T_1$ and $T_2$ will be approximated with a model of the form described in Section~\ref{subsec:model} as sketched in Figure \ref{fig:T=T_1+T_2}. 
Further, we define $H_2$ ($E_2$) to be the numbers of people that recovered (died) from $T_2$.

\begin{figure}
	\resizebox{\linewidth}{!}{\input{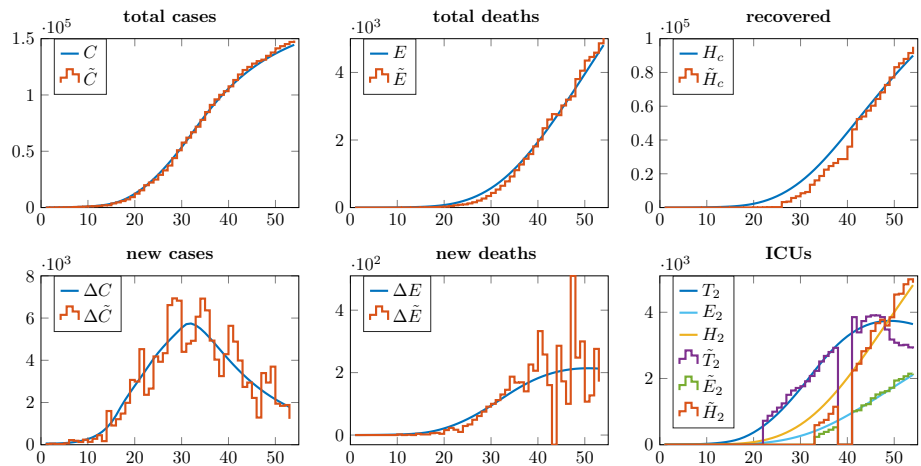}}
	\caption{Simulation of the model \eqref{eq:model} with the estimated parameters in Table \ref{tab:params} scaled by $N_\mathrm{total}$ compared to the actual data \cite{data_jhu}, \cite{divi2020report0404}. The horizontal axis represents the time in days, where $t=1$ is February 28 and $t=53$ is April 21.}\label{fig:fit}
\end{figure}

Finally, we perform the parameter optimization by solving a least squares problem via CasADi \cite{andersson2019casadi} to fit $C$, $E$, $H_c$, $T_2$, $E_2$, $H_2$ to the data $\tilde C$, $\tilde E$, $\tilde H_c$, $\tilde T_2$, $\tilde E_2$, $\tilde H_2$.
The best fitting parameters are given in Table \ref{tab:params} and the resulting fit is shown in Figure~\ref{fig:fit}.
Many of the constraints listed above are active at the optimal set of parameters, e.g., $\alpha=\gamma$, which is not surprising since we use the constraints to keep the parameters in a realistic range without further regularization.

This fit further enables us to specify the full model state of today $x(53)=:x_0$, which will be used in the following sections as the initial condition where $t=53$ corresponds to April 21
\begin{flalign}\label{eq:x_today}
x_0 &= \frac{1}{N_\text{total}} \begin{bmatrix}
82\,636\,256 & \!20\,581 & \!0 & \!8\,041 & \!41\,931 & \!11\,469 &\! 276\,911 & \!4\,810 \end{bmatrix}^\top\!\!\!\!.\hspace{-1cm}&
\end{flalign}
Please note that the model is quite sensitive to changes in the parameters and one obtains quite different parameter values if, e.g., the estimated range of unknown cases or the percentage of asymptomatic cases deviate from the assumptions.

\begin{table}
	\begin{tabular}{rlrlrlrlrl}
		\toprule
		$\alpha_\text{min}=$&$\mkern-15mu 0.0422$ & $u_1=$&$\mkern-15mu 0.5816$ &   $\zeta=$&$\mkern-15mu 0.0790$ &    $\mu_1=$&$\mkern-15mu 0.0080$ & $\mu_2=$    &$\mkern-15mu 0.0050$  \\
		$\alpha_\text{max}=$&$\mkern-15mu 0.3614$ & $u_2=$&$\mkern-15mu 0.7062$ &  $\lambda=$&$\mkern-15mu 0.0596$ & $\sigma_1=$&$\mkern-15mu 0.0370$ & $\sigma_2=$ &$\mkern-15mu 0.0552$  \\
		$\gamma_\text{min}=$&$\mkern-15mu 0.0422$ & $\beta=$&$\mkern-15mu 0.0084$ &   $\kappa=$&$\mkern-15mu 0.0563$ &   $\tau_1=$&$\mkern-15mu 0.0159$ & $\tau_2=$   &$\mkern-15mu 0.0242$  \\
		$\gamma_\text{max}=$&$\mkern-15mu 0.3614$ & $\theta=$&$\mkern-15mu 0.1981$ & $I_0=$&$\mkern-15mu 500/N_\mathrm{total}\mkern-40mu$ &   &$A_0=304/N_\mathrm{total}\mkern-100mu$ &            &  \\
		\bottomrule
	\end{tabular}
	\caption{The parameters of the model for Germany.}\label{tab:params}
\end{table}

\subsection{Modeling of the mortality rate}\label{subsec:mortality_rate}
It has been recognized as a key difficulty in handling the COVID-19 pandemic that the virus is highly contagious, thus infecting large numbers of individuals.
In addition, since many elderly and ill people require hospitalization and/or intensive care~\cite{verity2020estimates}, large waves of infections can quickly exceed the capacities of local health care systems~\cite{grasselli2020critical}.
Hence, ensuring that health care resources are sufficient is a key issue in handling the outbreak~\cite{IMHE2020forecasting}, given that an overwhelmed health care system even correlates positively with the mortality rate~\cite{ji2020potential}.

In this section, we describe how the mortality and recovery rates $\tau(T)$ and $\sigma(T)$ in~\eqref{eq:model} depend on the number of threatened patients $T$. 
The basic idea is that they are constant as long as the health care system's capacity is not at its limit, and the mortality rate $\tau$ (the recovery rate $\sigma$) increases (decreases) significantly if it is overwhelmed.
According to~\cite{divi2020report0404}, there are (on April 21) $2\,908$  COVID-19 patients in an ICU and $12\,623$ ICU spots are available.
Hence, the overall ICU capacity currently available for COVID-19 patients is $2\,908+12\,623=15\,531$, and we define the relative ICU capacity as $T_{\text{ICU}}=\frac{15\,531}{N_\mathrm{total}}$, where $N_\mathrm{total}=8.3\cdot 10^7$.
We consider a constant value of $T_{\text{ICU}}$ for simplicity, although it is likely that it will further increase in the future.

We assume that the mortality rate increases if the number of individuals requiring treatment in an ICU exceeds $T_{\text{ICU}}$, i.e., if $T_2>T_{\text{ICU}}$, with $T_2$ as in~\eqref{fig:T=T_1+T_2}.
More precisely, we assume that if a patient requiring intensive care does not receive it, then the patient deceases (i.e., for such patients, the mortality rate increases and the recovery rate is zero).
According to data of deceased individuals from Italy, those who were not admitted to an ICU deceased in median within $4$ days~\cite{iss2020report}.
Hence, we model those individuals in $T_2$ which are not admitted to an ICU via decaying first order dynamics with a half-life period of $4$ days, i.e, the corresponding time constant $\tau_{\text{crit}}$ satisfies $e^{-4\tau_{\text{crit}}}=0.5$, thus leading to $\tau_{\text{crit}}=0.173$.

In a first approximation $T_2\approx\frac{\mu_2}{\mu}T$ and hence we only modify the mortality rate $\tau$ in case that $\frac{\mu_2}{\mu}T>T_{\text{ICU}}$. 
In the model~\eqref{eq:model}, $\tau(T)$ and $\sigma(T)$ only occur jointly with $T$, which leads us to the following formula for $\tau(T)T$ and $\sigma(T)T$:
\begin{subequations}\label{eq:sigma_tau_mod}
\begin{align}
\tau(T)T &= \frac{\mu_1}{\mu}\tau_1 T+\max\left\{\frac{\mu_2}{\mu}\tau_{2}T,\tau_{2}T_{\text{ICU}}+\tau_{\text{crit}}\left(\frac{\mu_2}{\mu} T-T_{\text{ICU}}\right)\right\},\\\label{eq:sigma_mod}
\sigma(T)T&=\frac{\mu_1}{\mu}\sigma_1 T+\sigma_{2}\min\left\{\frac{\mu_2}{\mu}T,T_{\text{ICU}}\right\}.
\end{align}
\end{subequations}
If $\frac{\mu_2}{\mu}T\leq T_{\text{ICU}}$, then~\eqref{eq:sigma_tau_mod} implies $\tau(T)T=(\frac{\mu_1}{\mu}\tau_1+\frac{\mu_2}{\mu}\tau_2 )T$ and $\sigma(T)T=(\frac{\mu_1}{\mu}\sigma_1+\frac{\mu_2}{\mu}\sigma_2 )T$, i.e., a simple lumped model is recovered as long as the ICU capacity is not exceeded. 
If however $\frac{\mu_2}{\mu}T> T_{\text{ICU}}$, then the mortality rate increases to $\tau_{\text{crit}}$ for those $\frac{\mu_2}{\mu}T-T_{\text{ICU}}$ patients which require intensive care but do not receive it.
Similarly, for this fraction, the recovery rate is set to zero implicitly in~\eqref{eq:sigma_mod}.
The individuals $T_1 = \frac{\mu_1}{\mu}T$ not receiving intensive care are not affected by this mechanism.

Clearly, the modified rates in~\eqref{eq:sigma_tau_mod} are just a simple approximation of the effect that the mortality rate increases if hospitals are overwhelmed.
This modification in the model is crucial when studying the effect of loosening quarantine measures and corresponding optimal policies, as done in the remainder of this paper.
Since (fortunately) the German health care system has not been overwhelmed to this date, there are no quantitative data to validate the above modification and in particular, the exact value of $\tau_{\text{crit}}$.
Nevertheless, the refinement is confirmed \emph{qualitatively} by experiences in other countries~\cite{grasselli2020critical,IMHE2020forecasting,ji2020potential}.
Moreover, even a substantial change of $\tau_{\text{crit}}$ has little effect on the overall dynamics since it only affects the exact number of fatalities.
In particular, changing $\tau_{\text{crit}}$ does not lead to a qualitative change in an optimal policy to control the outbreak as long as $\tau_{\text{crit}}$ is sufficiently larger than $\tau_2$ and it is possible not to exceed the ICU capacity.
%
\section{Open-loop optimal control of the COVID-19 outbreak}\label{sec:nom_oc}
In this section, we discuss different policies that can be considered to keep the number of fatalities due to COVID-19 low while at the same time also impose as little constraints as possible on the public. The most significant degree of freedom currently is certainly influencing the infection rates $\alpha$ and $\gamma$. Measures for influencing the infection rates include hygienic measures, face masks, and different nuances of distancing policies, up to a mandated lockdown. Therefore, we define $u$ as introduced in Section~\ref{subsec:parameters} as our input, representing distancing policies or other measures that have a direct influence on the infection rates $\alpha$ and $\gamma$. We model this influence via
\begin{subequations}
\label{eq:u_param}
\begin{align}
\alpha(t) &= \alpha_\text{max} + (\alpha_\text{min} - \alpha_\text{max}) u(t) \\
\gamma(t) &= \gamma_\text{max} + (\gamma_\text{min} - \gamma_\text{max}) u(t),
\end{align}
\end{subequations}
where a value of $u = 1$ hence represents the policies in Germany as of mid April (lockdown) and $u = 0$ represents no social distancing or other measures (i.e.\ corresponding to infection rates as in the beginning of March).  
Furthermore, we assume that the policies affecting the infection rates $\alpha$, $\gamma$ (i.e. $u$) stay constant for at least one week and can only be changed every seven days.
In the first subsection, we will introduce different baseline policies which can give insights into the effects of different inputs $u$ and which will serve as a comparison to the optimal controller in the following subsection. More specifically, these baseline strategies will be used to define an upper bound on the social distancing measures that the optimal control in Section~\ref{sec:OC_3} and later on the feedback strategies in Sections~\ref{subsec:roc_1} and~\ref{subsec:roc_2} can employ to minimize the fatalities.

\subsubsection*{Test capacity}
In addition to varying the infection rates $\alpha$ and $\gamma$, another degree of freedom to influence the model~\eqref{eq:model} lies in adapting the testing policy. Testing individuals on COVID-19 is represented in the current model by parameters $\theta$ and $\epsilon$ for symptomatic and asymptomatic individuals, respectively. In the following, we assume that only a fixed number of tests can be carried out every day.
If we wish to only test symptomatic individuals, this includes both symptomatic individuals infected with COVID-19 (i.e., members of the state $A$) and individuals suffering from other illnesses with similar symptoms. 
In \cite{RKI2017}, the Robert Koch Institute estimates numbers on influenza-like illnesses (ILI) in Germany. 
While the numbers show clear seasonal differences, approximately $1.3\%$ of the population become newly infected with ILI on average per week, and approximately $37\%$ of them see a doctor (an indication for more severe symptoms). Moreover, influenza symptoms usually last 4-5 days leaving us with an approximate average of $p_{\text{sick}}=0.3\%$ of the population showing significant influenza-like symptoms at an arbitrary time of the year. When testing asymptomatic individuals, this includes infected persons without symptoms but also any other individual not known to be infected or healed and not showing any symptoms (i.e. $S+I+H-H_c-p_{\text{sick}}$, where $H_c$ are the confirmed healed cases, compare Section~\ref{subsec:parameters}).
The total amount of resources used for testing is then captured by the following cost: 
\begin{align}
\label{eq:c_test}
c_{\text{test}}(\epsilon,\theta,A,S,I,H-H_c)=\epsilon (S+I+H-H_c-p_{\text{sick}}) +\theta \cdot (A+p_{\text{sick}})+A\mu. 
\end{align}
Denoting the parameter $\theta$ from Table~\ref{tab:params} as $\theta_n$, we assume a fixed bound $\overline{c}>0$ on the amount of resources for testing $c_{\text{test}}$ and that $\theta_n$ corresponds to the \emph{nominal} value for $A=0$, i.e., $\overline{c}:=c_{\text{test}}(0,\theta_n,0)$.
In the following, we assume that the current policy with respect to testing stays in place: all available tests are used on a daily basis for as many symptomatic people as tests are available. 
Then the testing policy used throughout this section reads $\epsilon(t)=0$ (as is current practice) and
\begin{align}
\label{eq:u2_baseline}
\theta(t)=(\theta_n\cdot p_{\text{sick}}-\mu A(t))/(p_{\text{sick}}+A(t)).
\end{align}
Note that this also implies that throughout this paper the state $D \equiv 0$.
The allocation of tests (with the possibility of also saving test resources for later) can also be modeled as control inputs. However, in the present model the effects of temporarily saving tests (under the current resource constraints) are negligible compared to the effects of changes in the infection rates. 
Increasing the overall test capacity or improving the choice of test subjects (e.g. with tracing of cases), which corresponds to increasing values of $\theta_n$/$\overline{c}$, on the other hand, can potentially improve the evolution of the pandemic significantly, since detected individuals are less contagious than undetected ones. However, increasing test capacities or better allocated testing (especially with regard to $\epsilon$, i.e.\ tracing of also asymptomatic infections) is at the current stage not included in our consideration but could be addressed in future work with the presented model by choosing $\epsilon \neq 0$ and making $\theta_n$/$\overline{c}$ an increasing and time-varying variable.  

\subsubsection*{Control goal}
Given the introduced control input $u$, different control goals can be formulated. One such goal could be to obtain herd immunity. Herd immunity corresponds to the only stable equilibrium given no social distancing measures (i.e. with $\alpha_\text{max}$, $\gamma_\text{max}$) and requires a large part of the society to be immune. More precisely, herd immunity is reached if $S<S^\star$, where \cite{Giordano2020} provide a formula for calculating $S^\star$ (see Section~\ref{sec:nom_oc_1_1} for more details). Given our model, we can now calculate the minimum time that is needed to reach herd immunity. For this, we assume that we can choose a policy that utilizes the full health care capacity at all times. With $T \equiv \frac{\mu}{\mu_2} T_{\text{ICU}}$, $\dot{T} \equiv 0$, $\dot{A} + \dot{R} \equiv 0$, $\dot{I}+\dot{D} \equiv 0$, $S$ decreases each day by $\frac{(\zeta + \lambda)(\mu + \kappa)(\sigma + \tau)}{\zeta \mu} \frac{\mu}{\mu_2} T_{\text{ICU}}$. Hence,
\begin{align*}
t_{\text{herd}} = \frac{\zeta \mu_2 (1-S^\star)}{(\zeta + \lambda)(\mu + \kappa)(\sigma + \tau) T_{\text{ICU}}}
\end{align*}
gives a lower bound on the time required to reach herd immunity without exceeding the health care capacity given the introduced model. The herein identified model parameters yield a time span $t_{\text{herd}}$ of more than six years. 
A stable steady state in the absence of a vaccine (i.e. herd immunity) can hence only be obtained either after many years or by overstraining the health care system and a corresponding significant rise in the number of fatalities. Therefore, our ongoing assumption throughout this section is that prior to herd immunity, a vaccine will be available and we assume the availability of the vaccine in approximately two years. Our goal is thus to find an optimal policy minimizing the number of fatalities for the next two years while imposing as little constraints as possible on the public and the economy. 

In the next subsection, we simulate and discuss the following policies: 
\begin{enumerate}
\item Keeping the social distancing measures in place (or even increasing the measures) until the virus is eradicated in Germany
\item Slowly (or more aggressively) loosening the distancing measures without overwhelming the health care capacities (while possibly risking a second wave). 
 \end{enumerate}
In fact, the presented baseline policies are similar to the policies suggested by the German "Helmholtz-Initiative" in \cite{Helmholtz2020}. 
We will discuss our conclusions in comparison to theirs at the end of the section. 

In Section~\ref{sec:OC_3}, we will then improve these baseline policies by applying optimal control techniques and we will discuss the
importance and significance of the improvements.

\subsection{Introducing different baseline policies}
\label{sec:nom_oc_1}
\subsubsection{Consistent lockdown}
\label{sec:nom_oc_1_1}
In the following, we argue that a consistent lockdown strategy necessitates strong lockdown measures over a long time horizon to fully eradicate the virus as otherwise, dropping the social distancing measures too early leads to a second outbreak wave. 

Based on the SIDARTHE model fitted to the German outbreak, described in Section \ref{subsec:model}, we simulate how long we would need to remain in lockdown and simply wait for the virus to disappear. We define the disappearance of the virus as follows: If --- most probably --- there is less than one active contagious case, i.e., $I+D+A+R+T<0.5/N_\mathrm{total}$, the virus is eradicated. It takes $305$ days, which is almost one year, until this condition is fulfilled and clearly the economical and psychological damage caused by a lockdown period this long is excessive such that staying and waiting in lockdown is not an option. With even stricter measures, such as $\alpha_3=0.8\alpha_3$, $\gamma_3=0.8\gamma_3$, we could only marginally accelerate this process to $288$ days while increasing social distancing is costly, cf. the cost function in Section \ref{sec:OC_3}. Note that the equilibrium attained under the above lockdown policy is an unstable one that is not robust to uncertainties. In particular, if only one person remains infected when the measures are suspended they could cause a new outbreak. Also, the virus may be reimported from other countries or humans might be reinfected by an interim host.

Next, we simulate the following three scenarios in all of which the German population is kept in lockdown for a predefined period of time, followed by no measures at all. The only difference is the length of the lockdown period. In the first scenario, the measures are abolished immediately (April 21). The second one keeps the current strict measures for an additional $50$ days. The third variant simulates an even longer lockdown period, ending after $150$ days counting from April 21.

\begin{figure}
	\includegraphics[width=\linewidth]{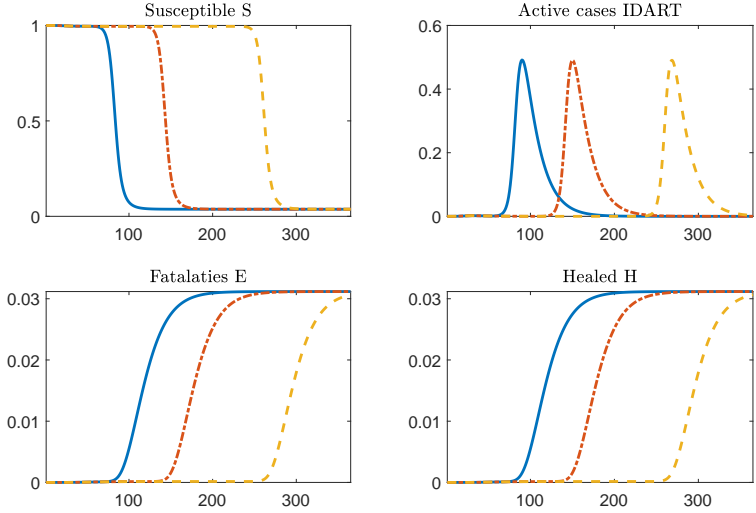}
	\caption{Simulation of the model \eqref{eq:model} for a consistent lockdown of different lengths $T_{l,i}$ with days on the horizontal axis ($T_{l,1}=0$ - blue, solid; $T_{l,2}=50$ - red, dash-dotted; $T_{l,3}=150$ - yellow, dashed).}\label{fig:consistent_lockdown}
\end{figure}

In Figure~\ref{fig:consistent_lockdown}, we compare the three scenarios.
We clearly see that in all three cases the number of currently infected people $I+D+A+R+T$ rises drastically a few days after the measures are removed independent of how long the lockdown persisted before. In any case, we experience a second outbreak wave. 
Staying longer in lockdown slightly delays the following peak of the share of active cases $I+D+A+R+T$, yet the peak amplitude is almost the same in all three scenarios.

This behavior can be explained as follows. If there is no one who currently has the virus, i.e., $I_{\textrm{eq}}=D_{\textrm{eq}}=A_{\textrm{eq}}=R_{\textrm{eq}}=T_{\textrm{eq}}=0$, such that $S_{\textrm{eq}}+H_{\textrm{eq}}+E_{\textrm{eq}}=1$, an equilibrium point is attained. The stability of the equilibrium point depends on the value of $S_{\textrm{eq}}$ and the model parameters.
In \cite{Giordano2020}, the authors show that the IDART subsystem is asymptotically stable if and only if $S_{\textrm{eq}}<S^\star$, where $S_{\textrm{eq}}$ is the susceptible state at equilibrium for a given initial condition $x_0$ and the corresponding parameters, especially for $\alpha$ and $\gamma$ that are actively adjusted according to an underlying policy. 
The value of $S^\star$ follows from the stability analysis of the linearized IDART subsystem and has the following structure with respect to $\alpha$ and $\gamma$
\begin{equation}
S^\star=\frac{a_1}{\alpha a_2+\gamma a_3+a_4},
\end{equation}
where $a_i, i=1\dots4$ are constants, see \cite{Giordano2020} for details and the definition of $S^\star$. Note further that the commonly stated base reproduction rate \eqref{eq:R_nought} is directly linked to the value $S^\star$ via $R_0=1/S^\star$. The stability of an equilibrium that depends on the parameters changes once we adjust $\alpha$ and $\gamma$. For strict measures ($\alpha_{\mathrm{min}}$, $\gamma_{\mathrm{min}}$), the value $S^\star$ is high ($S^\star= 2.242$), such that a stable equilibrium is attained for any $S$. This means that only a small number of people is infected by the virus before the equilibrium is attained. With no measures ($\alpha_{\mathrm{max}}$, $\gamma_{\mathrm{max}}$), a stable equilibrium requires the share of susceptible people to be smaller than $S^\star= 0.292$, i.e., herd immunity.

We can hence conclude that if $S_{\textrm{eq}}(x_0,\alpha_3,\gamma_3)>S^\star(\alpha_0,\gamma_0)$, the equilibrium attained during lockdown is unstable with no measures.
This means, there inevitably is a second outbreak wave once the lockdown ends. For the fitted model of the German outbreak, $S_{\textrm{eq}}(x_0,\alpha_3,\gamma_3)= 0.9956$ is attained after the first wave. Hence, at least another 70.4\% of the German population get infected in the second wave before a stable equilibrium is attained.
Altogether, this leaves us with the following conclusion of two possible outcomes when choosing a consistent lockdown strategy:
\begin{itemize}
\item Strong lockdown measures over long time horizons have to be taken to eradicate the virus in Germany. However, this takes a big toll on the public and any infected person, e.g. from abroad, could spark a second wave at any point.
\item Any lockdown strategy that does not fully wipe out the virus inevitably yields a second outbreak wave once all measures are suspended.
\end{itemize}

\subsubsection{Iterative loosening of the distancing policies}
\label{sec:nom_oc_1_2}
As argued in the previous subsection, keeping the lockdown policy strict can never lead to a stable equilibrium when ending the lockdown, no matter how long it did take place before all measures were suspended.
Hence, many countries are now discussing or have even already started to loosen the lockdown in very small steps. 
Indeed, experts consulting the German government ("Nationale Akademie der Wissenschaften Leopoldina") have recently published their recommendations concerning a possible strategy for loosening the lockdown gradually in small steps \cite{Leopoldina2020}. 

They name the following conditions for loosening the lockdown in small steps:
\begin{itemize}
\item[a)] The number of new infections remains at a low value.
\item[b)] The capacity of the health care system must not be exceeded.
\item[c)] Precautions (such as hygienic measures, face masks,  
distancing) remain in force.
\end{itemize}

In the following, we try to translate these recommendations into a policy for our simplified model to first analyze the results and second, to use this as a baseline policy for the optimizer in the following subsection. We implement the conditions presented above 
via the following policy strategy:
\begin{itemize}
\item[a)] $u$ can only be decreased if, over the last $n_{\text{stab}}$ days, the number of newly infected persons (i.e. $S(t-1) - S(t)$) is decreasing and $u$ has not been increased 
 \item[b)] $u$ can only be decreased if less than $X_{\text{lower}}$ of the ICUs are occupied
\item[c)] The decrease in $u$ can only be a small decrease at a time and therefore, the interval between $u_{\max} = 1$ (lockdown) and $u_{\min}=0$ (no measures) is divided into $n_{\text{steps}}$ equidistant steps.
\end{itemize}
Additionally, we add that $u$ will be increased again (with the same step size as the decrease)  if more than $X_{\text{upper}}$ of the ICUs are occupied and no decrease in the amount of necessary ICU is witnessed.
This policy results in four 'tuning parameters' of the policy: $n_{\text{stab}}$, $X_{\text{lower}}$, $X_{\text{upper}}$  and $n_{\text{steps}}$.  
In fact, it turns out that the outcome of the simulation is not very sensitive to the tuning parameters of the policy, but can 
be tuned to be slightly more careful or more aggressive. In the following subsection, we choose two different sets of parameters as baseline policies for the optimal control approach.

\subsection{Optimal control strategy}
\label{sec:OC_3}
In this section, we contrast the baseline policy from Section~\ref{sec:nom_oc_1_2} with an optimal control policy, under idealized assumptions (exact model and state measurement). 
The purpose of this section is twofold: 
a) Understand how an optimal policy differs qualitatively from the baseline policies. 
b) Quantify the loss of performance (in terms of increased fatalities and/or unnecessary social policy $u$) resulting from using a suboptimal baseline policy. 
The degree of freedom is the input $u\in[0,1]$ affecting the social policy and we consider the fact that the policy can only be changed every $T_s=7$ days.

\subsubsection*{Multi-objective optimal control problem}
In the following, we consider the problem only for the next $N=100$ weeks, assuming that thereafter a vaccine might be developed that would ideally prevent (almost all)  further fatalities in the future.
The overall control problem can be seen as a multi-objective optimal control problem, where we wish to simultaneously minimize the number of fatalities $E$
and  the societal and economical cost of the social policy measures, which will be measured by the function $c_{\text{policy}}(u)={1}/{\alpha(u)}$. 
We point out that due to the parametrization~\eqref{eq:u_param} this cost also inherently considers the infection rate $\gamma$.  
This cost function is such that the social cost of achieving an arbitrarily small infection rate $\alpha$ grows unbounded, while for large values of $u$ incremental differences are less relevant. 
In order to suitably characterize an \textit{optimal} solution to this multi-objective problem we use the baseline policies  in Section~\ref{sec:nom_oc_1}. 
The resulting optimal control problem is given by~\eqref{eq:OC} below, which will be explained in the following. 
In particular, our goal is to find an input policy that minimizes the number of accumulated fatalities, while using less resources than the baseline policy in terms of accumulated social impact of $c_{\text{policy}}$ (c.f.~\eqref{eq:OC_policy}).
We point out that similar ``stabilization'' problems subject to resource constraints for the control of epidemic outbreaks can be found in the literature, also using a fractional cost $c_{\text{policy}}$, compare e.g.~\cite{preciado2014optimal,kohler2018dynamic}. 

When minimizing the number of accumulated fatalities, it is important to consider not only the extinct individuals $E(N\cdot T_s)$ at the end of the two year horizon, but to account as well for the part of the already infected individuals that will decease \emph{after} the two year horizon.
The reason for this is, while the availability of a vaccine at the end of the horizon might prevent future infections, it cannot cure already infected people.
Hence, if we do not account for the inevitable fatalities among the individuals infected at the end of the prediction horizon, the optimal controller does not take any efforts to keep them low and as a result a lot of people would die shortly after the two year horizon.
Therefore, we propose an optimization objective that includes all past and inevitable future fatalities.
Based on the model~\eqref{eq:model}, we know that a total of $\frac \zeta {\zeta + \lambda}(I +D)$ of the infected people $I+D$ will develop symptoms in the future and further that a total of $\frac \mu {\mu + \kappa}\big(\frac \zeta {\zeta + \lambda}(I +D)+A +R\big)$ will become threatened. 
Thus, assuming the capacity $T_\text{ICU}$ is not exceeded afterwards, i.e., setting constant values $\tau=\tau(0)=\frac{\mu_1}{\mu}\tau_1+\frac{\mu_2}{\mu}\tau_2$ and $\sigma=\sigma(0)=\frac{\mu_1}{\mu}\sigma_1+\frac{\mu_2}{\mu}\sigma_2$, the amount of inevitable fatalities is exactly given by
\begin{align}\label{eq:terminal_cost_function}
F = E + \frac{\tau}{\tau+\sigma} \left(\frac \mu {\mu + \kappa}\left(\frac \zeta {\zeta + \lambda}(I +D)+A +R\right)+T\right).
\end{align}

Hence, given a baseline solution $u^b,x^b$ from Section~\ref{sec:nom_oc_1_2}, the corresponding optimal control problem reads as follows:
\begin{subequations}
	\label{eq:OC}
	\begin{align}
	\label{eq:OC_1}
	\min_{u(\cdot)} &~F(N\cdot T_s)\\
	\label{eq:OC_3}
	&0\leq u(k\cdot T_s)\in[0,1]\\
	& k=0,\dots N-1\\
	\label{eq:OC_policy}
	&\sum_{k=0}^{N-1}c_{\text{policy}}(u(k\cdot T_s))\leq \sum_{k=0}^{N-1}c_{\text{policy}}(u^b(k\cdot T_s)).
	\end{align}
\end{subequations}
Since we only change the policy every week, the index $k$ in~\eqref{eq:OC} corresponds to weeks and $F(N\cdot T_s)$ corresponds to the objective function in \eqref{eq:terminal_cost_function}, where the states result from simulating the system~\eqref{eq:model} with the parameters and the initial condition from Section~\ref{sec:modeling} and the input $u(\cdot)$ over $k$-weeks. 
Condition~\eqref{eq:OC_policy} ensures that the encountered social cost is smaller than the cost of the baseline policy.
This optimal control problem is such that the baseline policy $u=u^b$ is a feasible solution and thus the resulting fatalities $F(N\cdot T_s)$ will always be lower than that of the baseline policy. 
We point out that it is possible to consider a more restrictive transient constraint on the policy cost instead of~\eqref{eq:OC_policy}, which is discussed in detail in~\ref{app:OC}. 
The optimal control problem~\eqref{eq:OC} can be formulated as a nonlinear program (NLP) and is subsequently solved using CasADi~\cite{andersson2019casadi}.

\subsubsection*{Numerical results}
\begin{figure}
\begin{center}
\includegraphics[width=0.75\textwidth]{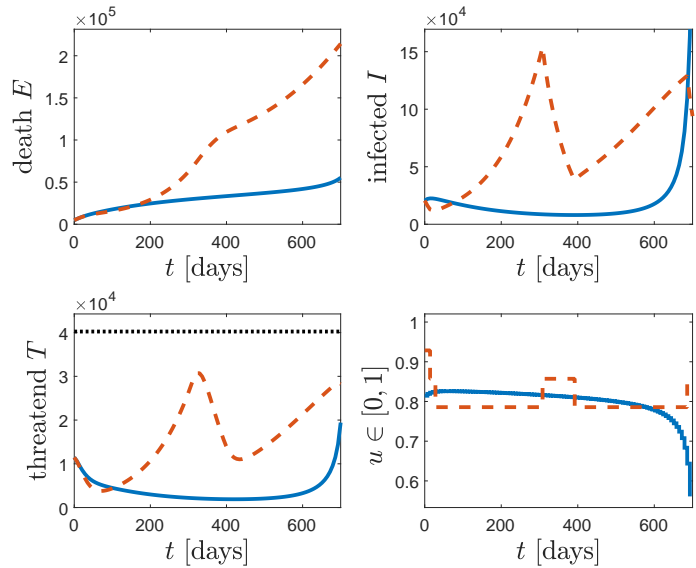} 
\caption{Optimal control strategy (blue, solid), baseline  policy (red, dashed) and ICU capacity (dotted, black).}  \label{fig:OC_1}
\end{center}
\end{figure}

For comparison and to implement the constraint~\eqref{eq:OC_policy}, we use the baseline policy from Section~\ref{sec:nom_oc_1_2} with $X_{lower}=0.4$, $X_{upper}=0.7$,
 $n_{\text{steps}}=14$, $n_{\text{stab}}=14$, which overall is rather cautious and does not exceed the ICU capacity. 
The corresponding results for the baseline policy and the optimal control strategy can be seen in Figure~\ref{fig:OC_1}. 
Although the optimal control input yields initially (first $100$ days) a slightly larger number of infected individuals and thus slightly more fatalities in the first 200 days, the number of infected individuals is subsequently significantly lower and the overall number of fatalities is reduced to only $26\%$. 
The optimal controller allows for a smooth increase of the infection rate $\alpha$, while keeping the number of threatened individuals ($T$)  consistently below the corresponding value of the baseline policy after the first 200 days, thus yielding a small number of fatalities.  The rising number of infected individuals ($I$) at the end results from the finite-horizon and will be considered later in more detail.

\begin{figure}
\begin{center}
\includegraphics[width=0.75\textwidth]{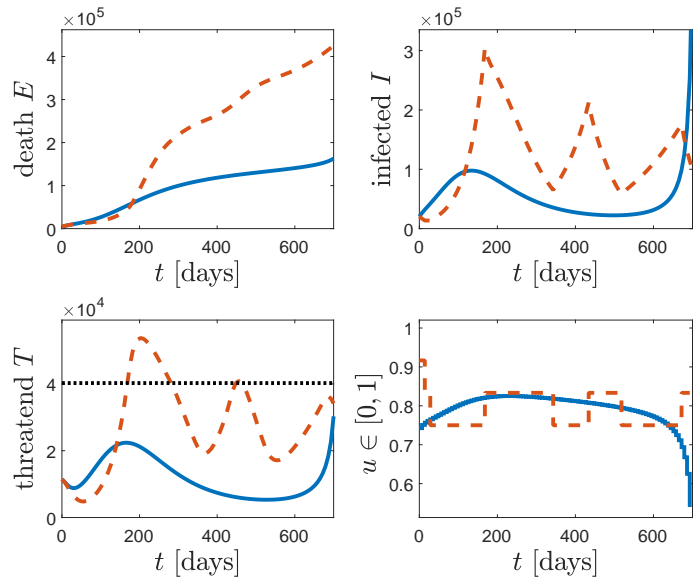} 
\caption{More aggressive baseline: Optimal control strategy (blue, solid), baseline policy (red, dashed) and ICU capacity (dotted, black).} \label{fig:OC_2}
\end{center}
\end{figure}

We also consider a second baseline policy using $X_{lower}=0.6$, $X_{upper}=0.85$, $n_{\text{steps}}=12$, $n_{\text{stab}}=14$, which slightly relaxes the social policy, but also exceeds the ICU capacity.
The result is shown in Figure~\ref{fig:OC_2}.
We can see that in comparison to this second baseline policy the optimal policy significantly reduces the number of fatalities to only  $39\%$. 
The optimal strategy is a lot more cautious in reducing the social policy, while the baseline is more aggressive and goes back and forth between increasing and decreasing $\alpha$, resulting in significant violations of the ICU capacity.
Furthermore, the simple baseline policy results in a second wave as the restrictions are loosened too quickly, while the optimal strategy slowly but steadily increases $\alpha$ after the first 200 days, and thereby avoids a second wave.

In both examples, a further observation should be highlighted.
After an initial phase of containing the outbreak, the measures are slowly but steadily relaxed until a larger release at the end of the horizon.
Similar behavior can be observed for many optimal control problems with finite horizons and is commonly referred to as ``turnpike'' behavior, which goes back to \cite{Dorfman1958}.
An explanation for this is that the consequences of decisions taken at later points in time mainly occur after the end of the horizon, such that a more aggressive policy towards the end is optimal, when only considering the finite two year horizon.
Of course, one would not implement the ``leaving arc'' if the development of a vaccine would not be finalized after two years, since implementing such a policy may lead to an uncontrollable increase of infections towards the end of the time horizon in case that the model is inaccurate and should thus be avoided in practice.
In~\ref{app:OC_term}, we therefore discuss how adding ``terminal constraints'' to the optimization problem can prevent this turnpike behavior of the optimal solution, at the price of an increasing number of fatalities.

\subsection{Discussion}
\label{sec:OC_4} 
If we compare the results in Section~\ref{sec:OC_3} with the consistent full lockdown from~Section~\ref{sec:nom_oc_1_1}, we can see that it is possible to appropriately increase $\alpha$ without exceeding the ICU capacities, while the consistent full lockdown strategy would require a lockdown that takes approximately a year to be effective. 
Hence, while a consistent lockdown can effectively minimize the number of deaths, this strategy is only viable in case this lockdown can be prolonged over the corresponding time horizon, unless a vaccine is developed earlier. On the other hand, both the optimal controller and the baseline controller allow for a significant relaxation of the lockdown (on average a doubling of $\alpha$), without significantly increasing the number of fatalities. 

In comparison to the baseline policy suggested in Section~\ref{sec:nom_oc_1_2}, the optimal control policy results in a slower but smooth loosening of the distancing policies.
Without increasing the social cost over the full time horizon, this optimal policy avoids any violation of the maximum ICU capacity and hence results in a significantly smaller fatality rate. We point out that the resulting optimal policy of slowly increasing $\alpha$ is qualitatively similar to the resulting optimal policy in~\cite{kantner2020beyond}, albeit for a different control goal.
It can be seen that an initially ``stronger'' lockdown (i.e., a smaller value of $\alpha$) over a longer time period with subsequent loosening leads to a better handling of the pandemic, compared to repeated tightening and loosening of distancing measures.
Moreover, a smooth and monotone loosening of distancing policies is also desirable from an economic aspect since repeated lockdowns after interim-periods of relaxed distancing guidelines may be even more damaging to the economy, compared to an initially longer lockdown.

Comparing our results with the proposed scenarios by the Helmholtz Association \cite{Helmholtz2020}, we find that we agree that the goal of herd immunity without overwhelming the health care capacity would require years. Our results further agree with \cite{Helmholtz2020} that the contact restrictions can only be loosened slowly if the health care capacity must not be overwhelmed. However, since the authors in \cite{Helmholtz2020} do not consider the availability of a vaccine, their conclusion is keeping or even increasing the lockdown until the number of infected persons is small and all infections can be traced efficiently and effectively via strategically allocated (and increased) testing. With the assumption of a vaccine within the next two years, we argue that a slow and smooth loosening of the lockdown does not lead to many more fatalities while decreasing the social and economic cost significantly according to our model. Concurrently, increased and strategically better allocated testing, e.g., via a contact tracing mobile app \cite{Oliver2020}, is of course highly beneficial and greatly advisable to improve the performance (even if it this was not accounted for in our model).  

To summarize, it seems possible to reduce the current restrictions and thus allow $\alpha$ to increase without exceeding the ICU capacity. 
Furthermore, optimized policies can significantly improve the outcome (in terms of fewer deaths and/or less social restrictions).  
However, the result is highly sensitive w.r.t. the change in the infection rate, while an accurate control of the infection rate $\alpha$ (e.g. $\pm 5\%$) through governmental policies seems difficult/unrealistic. 
In the next section, we will therefore deal with these issues by formulating a \textit{robust} control strategy that takes uncertainty in our COVID-19 model into account and uses feedback based on uncertain state information. 

%
\section{Optimal feedback control of the COVID-19 outbreak}\label{subsec:roc_1}

Section~\ref{sec:OC_3} shows that an optimal control policy can significantly reduce the number of fatalities compared to a baseline policy that allows for iterative loosening of social distancing measures.
This optimal control policy is computed by optimizing over all possible policies to find the one minimizing the number of fatalities predicted by the model equations~\eqref{eq:model} without using stronger shutdown measures than the baseline.
Hence, the policy proposed in Section~\ref{sec:OC_3} strongly relies on the accuracy of the model identified in Section~\ref{sec:modeling} and thus may fail to effectively control the outbreak in case of a model mismatch.
However, such a model mismatch is inevitable in practice, especially since the model itself is a simplification of a much more complex reality and the identification outlined in Section~\ref{subsec:parameters} strongly depends on the (sparse) available data and the additional prior knowledge based e.g. on further studies concerning COVID-19, which also provide only estimates.
In addition, the optimal control policy relies on exact knowledge of all states and on the assumption that values for $\alpha$ and $\gamma$ can be exactly imposed up to arbitrary precision via social distancing measures, both of which are unrealistic assumptions when applying the policy in practice.

In this section, we show how \emph{online measurements} can be utilized via \emph{feedback} to effectively and robustly control the German COVID-19 outbreak in the presence of uncertain parameters.
More precisely, we illustrate that the optimal open-loop policy of Section~\ref{sec:OC_3} may lead to poor performance when applied to validation models with a different set of parameters, although these validation models result from adjusting only one prior assumption in the identification and still fit the past data well.
On the other hand, we show that a model predictive control (MPC) feedback strategy, based on repeatedly computing an open-loop policy for the nominal model from Section~\ref{sec:modeling}, is inherently robust w.r.t. model inaccuracies and successfully handles the outbreak.

\subsubsection*{Basic idea}
At each time step $k=1,\dots,N$, where $k$ corresponds to weeks and $N=100$ as in Section~\ref{sec:OC_3}, we solve the optimization problem \eqref{eq:OC} over the time horizon $k,\dots,N$ using the current measurements as initial condition at week $k$.
Then, we apply the computed optimal policy over one week before solving the problem for the new measurements again.
In this way, since the initial conditions in the optimal control problem are updated, a feedback mechanism is included as is standard in MPC~\cite{rawlings2017model}.
As a result, the prediction horizon $N-k$ of the optimization problem is shrinking with each time step $k$, such that it never exceeds the considered total time horizon of $N=100$ weeks, after which we assume the availability of a vaccine.
Hence, since the constraint~\eqref{eq:OC_policy} needs to hold over the whole time horizon $N$, we replace it by
\begin{align}
\sum_{j=k}^{N-1} c_{\text{policy}}(u(j\cdot T_s))\leq c_{\text{policy}}^1-\sum_{j=0}^{k-1} c_{\text{policy}}(u(j\cdot T_s)),
\end{align}
with $c_{\text{policy}}^1=\sum_{k=0}^{N-1}c_{\text{policy}}(u^b(k\cdot T_s))$ being the cost of the first baseline policy in Section~\ref{sec:nom_oc_1_2}.
As a second modification, we adapt the bound on the social distancing cost online, depending on the predicted states, as is detailed in the following.

\subsubsection*{Online adaptation of social policy constraint}
In Section~\ref{sec:OC_3}, we proposed an open-loop optimal control strategy, where the inputs were the infection rates $\alpha$ and $\gamma$.
Loosely speaking, the control goal was to achieve a minimum number of fatalities without imposing stronger social distancing measures than a simple baseline policy (compare~\eqref{eq:OC_policy}).
Since this constraint heavily depends on the model to which the baseline is applied, a realistic setting with imperfect model knowledge should allow to adapt the constraints on the policy online in case that the nominal model is overly optimistic or pessimistic.
Instead of simply requiring that the cost of the MPC-based feedback cannot exceed $c_{\text{policy}}^1$, we increase the maximum cost in case that the predicted number of patients requiring intensive care lies above $90\%$ of the maximum capacity $T_{\text{ICU}}$ at least once during the horizon, and we decrease it in case that the number consistently lies below $10\%$ of $T_{\text{ICU}}$.
This adaption is natural, as in reality one would increase the efforts to contain the outbreak if the current measures are insufficient, and on the other hand, the population cannot be expected to accept strict measures when there are only few (severe) cases across the country.
Therefore, the maximum cost in week $k$, denoted by $c_b(k)$, varies online with $k$ and is initialized as $c_b(0)=c_{\text{policy}}^1$.
The amount by which we change $c_b$ online is $\pm\Delta_u\frac{N-k}{N}$, where $\Delta_u=\frac{1}{\alpha_{\min}}-\frac{1}{\alpha_{\max}}$ with $\alpha_{\min}$ and $\alpha_{\max}$ as in Section~\ref{subsec:parameters}.
If, for instance, the model predicts large numbers of future ICU patients, then the cost bound $c_b$ is increased by the difference between the minimum and the maximum social distancing cost, scaled by the remaining time horizon.
This increase corresponds to the social cost of an additional week in full lockdown, scaled by the remaining time horizon via the factor $\frac{N-k}{N}$.

\subsubsection*{MPC-based feedback strategy}
The proposed MPC-based feedback strategy is summarized in Algorithm~\ref{alg:nominal_MPC}.
In the algorithm, $T(j\cdot T_s\mid k\cdot T_s)$ denotes the number of threatened individuals at time $j\cdot T_s$, predicted by the optimal solution of~\eqref{eq:OC} at time $k\cdot T_s$.
Essentially, the algorithm repeatedly applies the open-loop optimal control policy of Section~\ref{sec:OC_3} with the key difference that, at time $k$, all past measurements $j=1,\dots,k$ are used in the optimization problem, thus including an online feedback.
In addition, in Step 3 of the algorithm, the social policy constraints is adapted as described above.

\begin{algorithm}
\begin{Algorithm}\label{alg:nominal_MPC}
\normalfont{\textbf{MPC-based feedback strategy}}
\begin{enumerate}
\item Given state measurements up to time $k$, solve the following problem
\begin{subequations}
\label{eq:OC_fb}
\begin{align}
\label{eq:OC_fb_1}
\min_{u(\cdot)} &~F(N\cdot T_s)\\
\label{eq:OC_fb_3}
&0\leq u(j\cdot T_s)\in[0,1]\\
& j=k,\dots,N-1\\
\label{eq:OC_fb_policy}
&\sum_{j=k}^{N-1} c_{\text{policy}}(u(j\cdot T_s))\leq c_b(k)-\sum_{j=0}^{k-1} c_{\text{policy}}(u(j\cdot T_s)),
\end{align}
\end{subequations}
with the state-dependent cost $F$ as in~\eqref{eq:terminal_cost_function}, based on simulating the model~\eqref{eq:model} over the remaining horizon $N-k$ subject to the input $u$, starting at the current measured state at time $k$.
\item Apply the optimal policy $u^*(k\cdot T_s)$ for the next $T_s=7$ days.
\item Update the social policy cost as
\begin{align*}
c_b(k+1)=\begin{cases}
      c_b(k)+\Delta_u\frac{N-k}{N}& \text{if $\frac{\mu_2}{\mu}\max_{j\in[k,N]}T(j\cdot T_s\mid k\cdot T_s)\geq0.9 T_{\text{ICU}}$}\\
      c_b(k)-\Delta_u\frac{N-k}{N}& \text{if $\frac{\mu_2}{\mu}\max_{j\in[k,N]}T(j\cdot T_s\mid k\cdot T_s)\leq0.1 T_{\text{ICU}}$}\\
      c_b(k) & \text{otherwise}
    \end{cases}
\end{align*}
where $\Delta_u=\frac{1}{\alpha_{\min}}-\frac{1}{\alpha_{\max}}$.
\item Set $k=k+1$.
\end{enumerate}
\end{Algorithm}
\end{algorithm}

\subsubsection*{Validation}

To assess the improved robustness of Algorithm~\ref{alg:nominal_MPC} compared to open-loop optimal control, we produce two validation models.
More precisely, we identify two new sets of parameters A and B by proceeding exactly as in Section~\ref{subsec:parameters} with the only difference that we change the prior assumption that the stationary ratio of confirmed COVID-19 cases is in the interval $[0.3, 0.45]$.
Instead, we assume that this value is in $[0.3, 0.6]$ for set A and in $[0.3, 0.4]$ for set B.
When performing parameter identification based on these modified prior assumptions, we also obtain sets of parameters that can accurately explain the existing past data on COVID-19 cases in Germany.
However, the resulting models have different dynamics and different reproduction rates for the same lockdown policy.
Increasing the above ratio as in parameter set A decreases the number of infected and undetected individuals resulting in a higher reproduction rate to explain the same amount of confirmed cases.
Hence, if an open-loop policy based on the nominal model (i.e., with parameters described in Section~\ref{subsec:parameters}) is applied to the validation model with parameters A, then the number of infections and thus the number of fatalities increases significantly.

\begin{figure}
\begin{center}
\begin{minipage}{0.1\textwidth}
\textbf{A}\end{minipage} 
\begin{minipage}{0.75\textwidth}
\includegraphics[width=1\textwidth]{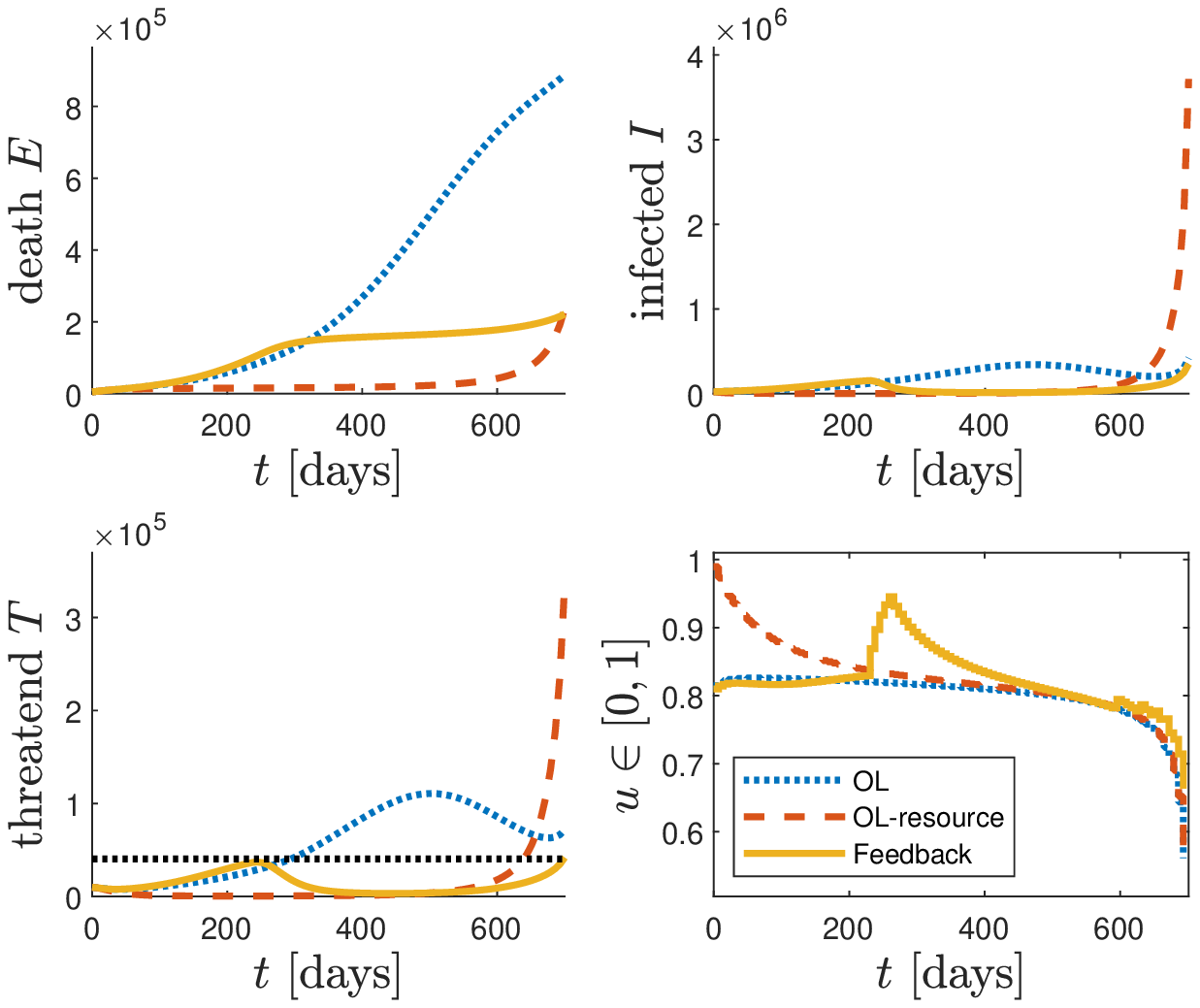}
\end{minipage}\\\vspace{6mm}
\begin{minipage}{0.1\textwidth}
\textbf{B}\end{minipage} 
\begin{minipage}{0.75\textwidth}
\includegraphics[width=\textwidth]{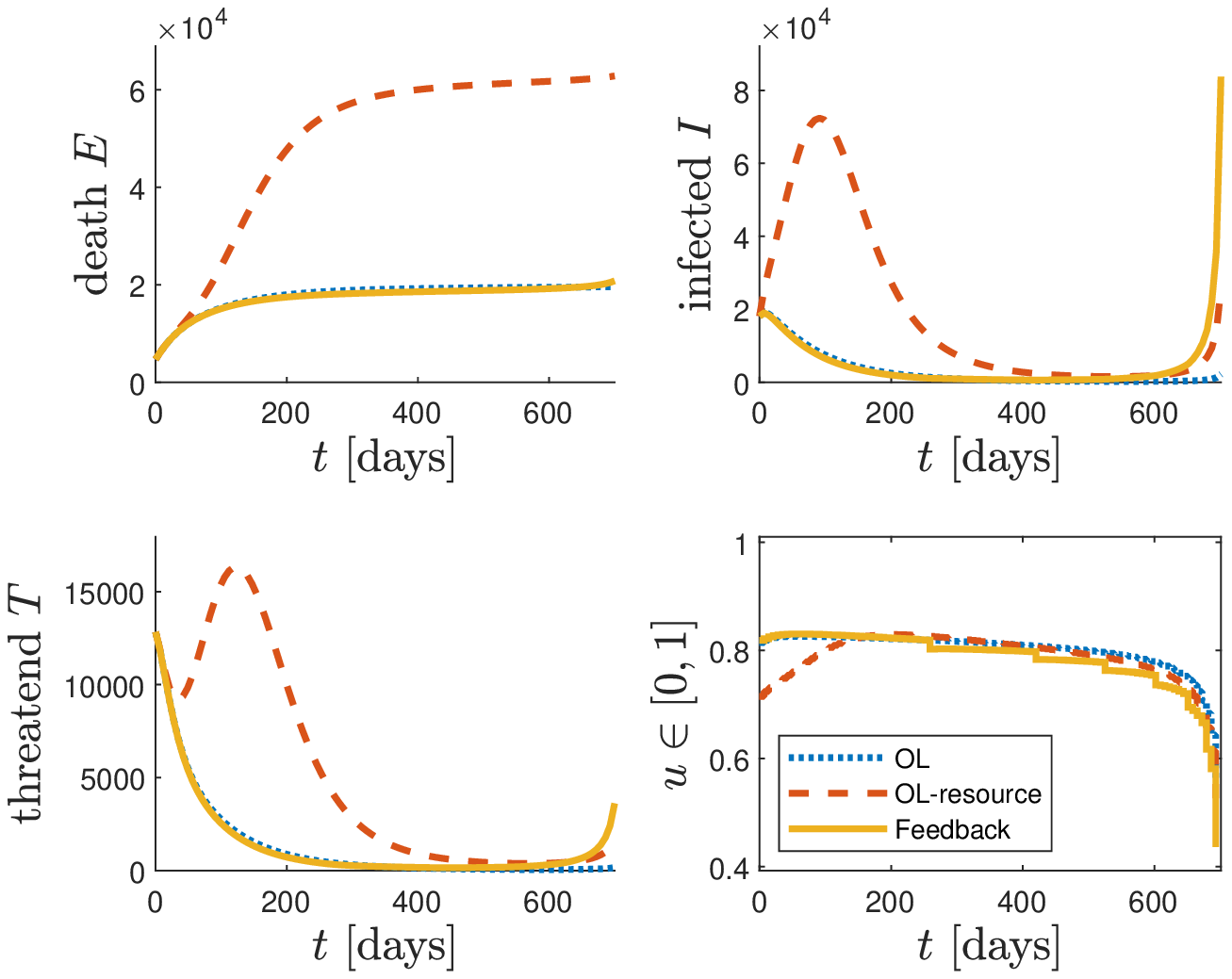} 
\end{minipage}
\caption{Different policies designed based on the nominal model identified in Section~\ref{subsec:parameters}, applied to the validation model with parameter set A (top) and parameter set B (bottom):
Open-loop optimal control strategy ('OL', blue dotted), Open-loop optimal control strategy with the same amount of resources as the MPC-based feedback  ('OL-resource', red dashed), MPC-based feedback strategy as in Algorithm~\ref{alg:nominal_MPC} (yellow solid), and ICU capacity (black dotted).} \label{fig:validation}
\end{center}
\end{figure}

To illustrate this effect, we apply the open-loop optimal control policy based on the model with parameters as in Section~\ref{subsec:parameters} to the new models with parameter sets A and B.
The control effort of this policy, i.e., the amount of social distancing, is constrained as in~\eqref{eq:OC_policy} by the cost of the baseline policy when applied to the nominal model identified in Section~\ref{subsec:parameters}.
The results for the model with parameters A can be seen in Figure~\ref{fig:validation}.
Since this validation model has a higher reproduction rate for similar inputs as explained above, the number of fatalities after $N=100$ weeks increases significantly compared to the simulations in Section~\ref{sec:nom_oc}.
This is due to the fact that the open-loop policy is only computed once, at the beginning of the time horizon, and is then applied over the whole time span of two years without any online adaption based on new measurements.
Therefore, it cannot handle the model mismatch and thus has a significantly worse performance.
In addition, Figure~\ref{fig:validation} shows the evolution under the proposed MPC-based feedback, which leads to a significantly lower number of fatalities compared to the open-loop policy.
We point out that the feedback (partially) compensates the fact that the control action is computed based on the nominal model parameters from Section~\ref{subsec:parameters}, which differ significantly from parameter set A.
Due to the larger number of infected individuals, the maximum social cost $c_b$ is increased at multiple time steps, which is indicated by the step-like increases of the input.
Finally, an open-loop optimal control policy is computed which is allowed to use the same amount of resources as the MPC-based feedback (in hindsight), i.e., the adapted social and economical cost $c_b(N)$.
While this policy performs better than the initial open-loop policy with fewer resources, it leads to a similar number of fatalities at the end of the horizon compared to the feedback controller.
However, the number of threatened patients is very large at time $k=N$, which would lead to a significant increase in fatalities after the considered time period, even if a vaccine is available.

Figure~\ref{fig:validation} also shows the same comparison for the model with parameter set B.
In this case, since the reproduction rate is lower, the open-loop optimal policy leads to fewer fatalities than in Section~\ref{sec:OC_3}.
The MPC-based feedback leads to almost identical performance, but it reduces the cost budget at several time instants, i.e., it can handle the outbreak similarly well but with significantly lower social and economic costs.
When restricting the budget of the open-loop optimal policy to the one of the feedback strategy, i.e., $c_b(N)$, then it leads to a dramatic increase in fatalities.

To conclude, the above discussion reveals that a combination of open-loop optimal control with feedback is inherently robust in the sense that it effectively controls the German COVID-19 outbreak even if the employed model is inaccurate.
When comparing the result to an open-loop strategy, then the MPC-based feedback strategy can dramatically decrease the number of fatalities or the necessary amount of social distancing, respectively.
Such robustness is an important property for applying any control strategy in a real-world scenario, where accurate model knowledge is rarely available.
In the next section, we propose a more systematic robust MPC approach which explicitly takes model inaccuracies as well as uncertain state measurements and control inputs into account in order to safely and cautiously control the COVID-19 pandemic.
\section{Robust and optimal feedback control of the COVID-19 outbreak}\label{subsec:roc_2}
While the MPC-based feedback policy proposed in Section~\ref{subsec:roc_1} is significantly more successful in handling the outbreak compared to a simple open-loop policy, it relies on the assumption that exact measurements of the state in~\eqref{eq:model} are available at each time step.
In this section, we consider a more realistic scenario of uncertain measurements in terms of biased state estimates, and we analyze the impact on the closed-loop operation.
In particular, we develop a robust MPC-based feedback strategy using interval arithmetic that takes the uncertainty into account during the predictions and thus leads to a safer policy minimizing the number of fatalities.
\subsubsection*{Biased state measurements}
In the following, we consider the case where at each day $k$ instead of the true state $x(k)$ we only obtain an estimated state $\hat{x}(k)$, which is subject to an additional bias.
In Table~\ref{tab:uncertain_states}, we summarize the uncertainties in the states. 
For individuals in states $D$ and $R$, the disease COVID-19 was detected by tests.
Hence, their values are well known, nevertheless, we assume that they can slightly differ from the true states by $\pm 1\%$, as there might be cases on the borderline between $D$ and $R$ that are hard to assign to either of the states. 
The number of people in ICUs is well documented. However, the state $T$ contains not only patients in ICUs ($T_{ICU}\pm1\%$) but also other infected members of the risk group ($T_2\pm5\%$), cf. Section \ref{subsec:model}, such that the uncertainty we use is $\pm(\frac{\mu_1}{\mu}\cdot1\%+\frac{\mu_2}{\mu}\cdot5\%)$. We assume that the number of deaths is certain by $\pm1\%$ as it includes some people that died of different causes. 
As the undetected cases can by definition not be measured, they must be estimated using random sampling or strategies like~ \cite{Bommer2020}. Therefore, the states $I$ and $A$ are much less certain, especially  without symptoms ($I\pm50\%$, $A\pm20\%$).
Recovering from the disease is a resulting state from both rather certain states, $D$ and $R$, and highly uncertain states, $I$ and $A$ such that overall it is uncertain itself ($H\pm50\%$). The uncertainty of the state of susceptible persons $S$ results from the other states: $\hat{S}=\hat{x}_1=1-\sum_{i=2}^8 \hat{x}_i$. 
\begin{table}
	\begin{tabular}{llllll}
		\toprule
		$I\pm50\%$ & $D\pm1\%$ & $A\pm20\%$ & $R\pm1\%$ & $H\pm50\%$ & $E\pm1\%$\\
		\multicolumn{6}{l}{$T\pm(\frac{\mu_1}{\mu}\cdot1\%+\frac{\mu_2}{\mu}\cdot5\%)$} \\
		\bottomrule
	\end{tabular}
	\caption{Uncertainties in the states.}\label{tab:uncertain_states}
\end{table}

It is possible to directly use this biased state estimate $\hat{x}(k)$ in Algorithm~\ref{alg:nominal_MPC} and compensate the bias through the inherent robustness in the feedback implementation.
In the following, we derive an alternative \textit{robust} formulation that explicitly considers the uncertainty in the prediction.

\subsubsection*{Interval predictions}
First, given a biased state estimate $\hat{x}(k)$ and known bounds on the bias (Tab.~\ref{tab:uncertain_states}), it is possible to compute interval bounds $\underline{x}(k)$, $\overline{x}(k)$ such that the true state is guaranteed to lie in that interval, i.e., $x_i(k)\in[\underline{x}_i(k),\overline{x}_i(k)]$.
The following formulation will predict the interval bounds $\underline{x}_i$ and $\overline{x}_i$ instead of using some nominal prediction. 
This methods is similar to interval arithmetic employed in  robust MPC~\cite{limon2005robust} and  the robust moment enclosure for an SEIV epidemic model in~\cite{watkins2019robust}. 
Using the fact that the system is positive ($x_i$ and all the parameters are positive), it is possible to derive an interval prediction of the form
\begin{subequations}
\label{eq:interval_predict}
\begin{align}
\dot{\overline{x}}=&\overline{f}(\underline{x},\overline{x},u),\\
\dot{\underline{x}}=&\underline{f}(\underline{x},\overline{x},u),
\end{align}
\end{subequations}
which ensures that $x(0)\in[\underline{x}(0),\overline{x}(0)]$ implies $x(t)\in[\underline{x}(t),\overline{x}(t)]$ for all $t\geq 0$, given suitable bounds on the uncertain parameters in the system model~\eqref{eq:model}. 
The detailed derivation of the interval prediction model~\eqref{eq:interval_predict} can be found in~\ref{app:interval} (more precisely, Equations~\eqref{eq:model_interval}).
Since deriving reliable bounds on all parameters in the model~\eqref{eq:model} is rather difficult or unnecessarily complex, we only focus on the uncertainty associated with the infection rate $\alpha$.
In particular, we consider an uncertainty of $\pm 5\%$ on the infection rate $\alpha$.
Thereby, we explicitly consider the problem that the infection rate cannot be precisely specified via social distancing measures.
We will see later in the simulations that, although we do not account for all possible mismatches in the prediction model, we nevertheless obtain the desired properties in closed loop. 

Given this interval prediction model, the proposed robust formulation now predicts an interval for the different state variables and minimizes the worst-case number of fatalities $\overline{F}$ based on $\overline{x}(N\cdot T_s)$. 
The overall procedure is summarized in Algorithm~\ref{alg:robust_MPC}.

\begin{algorithm}
\begin{Algorithm}\label{alg:robust_MPC}
\normalfont{\textbf{Robust MPC strategy using interval arithmetic}}
\begin{enumerate}
\item Given biased state estimate $\hat{x}(k\cdot T_s)$, compute set $[\underline{x}(k\cdot T_s),\overline{x}(k\cdot T_s)]$.
\item Solve the following problem
\begin{subequations}
\label{eq:OC_robust}
\begin{align}
\min_{u(\cdot)} &~\overline{F}(N\cdot T_s)\\
&0\leq u(j\cdot T_s)\in[0,1]\\
& j=k,\dots N-1\\
&\sum_{j=k}^{N-1} c_{\text{policy}}(u(j\cdot T_s))\leq c_b(k)-\sum_{j=0}^{k-1} c_{\text{policy}}(u(j\cdot T_s)),
\end{align}
\end{subequations}
with $\overline{F}$ based on~\eqref{eq:terminal_cost_function} using $\overline{x}$, which results from the interval predictions of the model~\eqref{eq:interval_predict} over the remaining horizon $N-k$ subject to the input $u$, starting at the current set estimate $[\underline{x}(k\cdot T_s),\overline{x}(k\cdot T_s)]$.
\item Apply the optimal policy $u^*(k\cdot T_s)$ for the next $T_s=7$ days.
\item Update the social policy cost as in Algorithm~\ref{alg:nominal_MPC} using $\overline{T}$ instead of $T$.
\item Set $k=k+1$.
\end{enumerate}
\end{Algorithm}
\end{algorithm}

\subsubsection*{Numerical results}
\begin{figure}
\begin{center}
\begin{minipage}{0.1\textwidth}
\textbf{A}\end{minipage} 
\begin{minipage}{0.75\textwidth}
\includegraphics[width=1\textwidth]{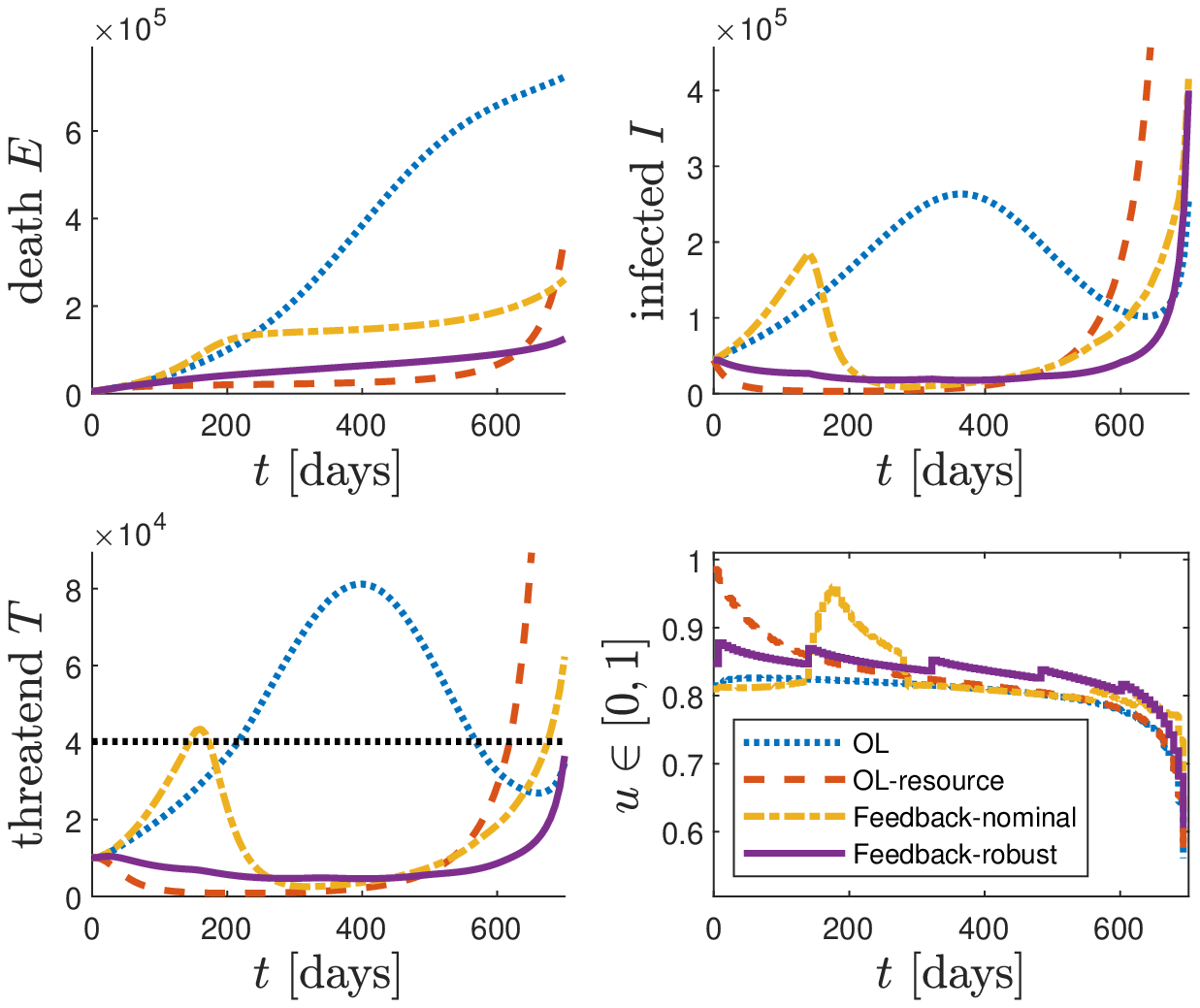}
\end{minipage}\\\vspace{6mm}
\begin{minipage}{0.1\textwidth}
\textbf{B}\end{minipage} 
\begin{minipage}{0.75\textwidth}
\includegraphics[width=\textwidth]{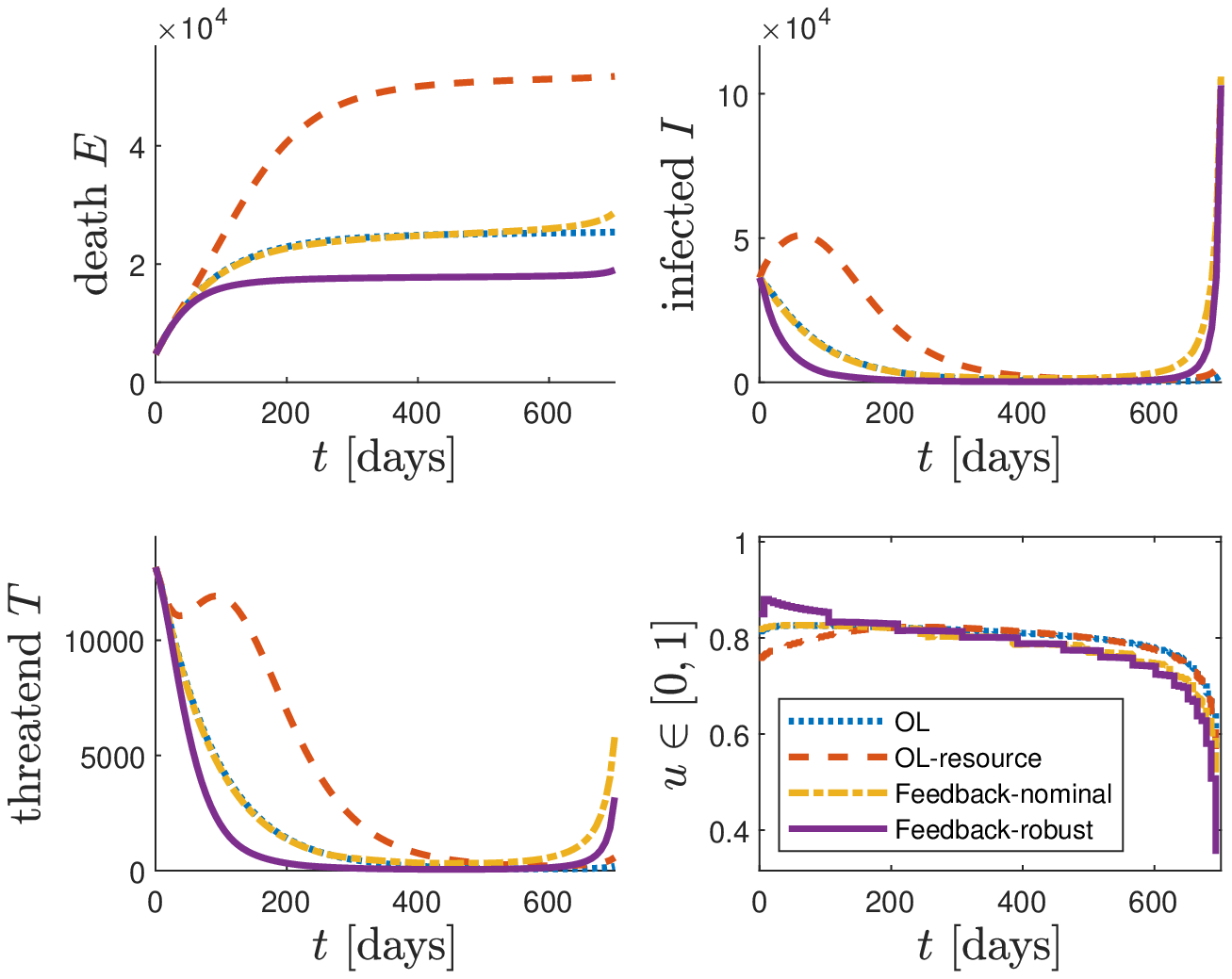} 
\end{minipage}
\caption{Different policies designed based on the nominal model identified in Section~\ref{subsec:parameters}, applied to the validation model with parameter set A (top) and parameter set B (bottom) with biased state measurements:
Open-loop optimal control strategy ('OL', blue dotted), Open-loop optimal control strategy with the same amount of resources as the robust MPC-based feedback  ('OL-resource', red dashed), nominal MPC-based feedback strategy as in Algorithm~\ref{alg:nominal_MPC} with biased estimated state $\hat{x}(k)$ (yellow, dash-dot), robust MPC-based feedback strategy using Algorithm~\ref{alg:robust_MPC} (purple, solid), and ICU capacity (black dotted).} \label{fig:tube}
\end{center}
\end{figure}

In the following simulations, we consider the extreme case where the number of estimated infected or previously infected individuals ($I$, $D$, $A$, $R$, $T$, $H$, $E$) is underestimated. 
The results for the robust MPC and the nominal MPC in comparison to open-loop optimal control strategies for the two validation parameter sets A and B (compare Section~\ref{subsec:roc_1}) can be seen in Figure~\ref{fig:tube}. 
Due to the worst-case prediction in the robust MPC, at $t=0$ the robust MPC already increases the resources two times, for both parameter sets, such that at $t=T_s$ the predictions satisfy $\frac{\mu_1}{\mu_2}\overline{T}(k\cdot T_s)\leq 0.9 T_{ICU}$. 

In the simulation with the model based on parameter set A, we can directly see that both the nominal MPC and the robust MPC reduce the number of fatalities compared to an open-loop optimal control strategy. 
Furthermore, if we compare the robust MPC and the nominal MPC, we can see that after $t=140$ days the nominal MPC implementation realizes that the spread is worse than initially assumed.
This leads to a strong increase in social measures $u$ at $t=140$. 
With the robust formulation, $u$ decreases almost monotonically.
Furthermore, the nominal implementation results in twice the number of fatalities, while the applied resources $c_{\text{policy}}$  over the two year horizon differ by less than $\Delta_u$, which corresponds to one week of lockdown. 
This indicates that the robust MPC, planning cautiously from the beginning, exploits its resources more efficiently by imposing stricter social distancing measures early on, which results to be beneficial in the long run.

For the second parameter set B, we can see that the worst-case robust formulation uses initially an unnecessarily high control effort. Nevertheless, overall the applied resources $c_{\text{policy}}$ differ by less than $\Delta_u$ compared to the nominal MPC, while the number of fatalities is still reduced by $33\%$. 
In comparison to the open-loop policies, both MPC policies require less or equally restrictive policy measures $u$, while the number of fatalities are significantly reduced.

\section{Conclusions and Discussion}
\label{sec:conclusions}
In the following, we summarize our findings on a high-level and highlight the main take-away messages:
\begin{itemize}
\item Our results in Section~\ref{sec:nom_oc} confirm the conclusions in \cite{Hermann2020} that neither eradication of the virus nor herd immunity without the availability of a vaccine are viable solutions to handle the current COVID-19 outbreak.
\item Applying an optimizer to the mathematical model describing the outbreak, one can significantly reduce the number of fatalities without increasing the costs associated to decreasing the infection rate (social distancing policies, closing schools, etc.), compare Section~\ref{sec:OC_3}. 
\item Since the proposed model can never exactly predict the COVID-19 pandemic, applying a nominal optimal policy introduces unnecessary conservatism, at best, up to posing a great danger (i.e.\ overwhelming the health care capacities risking high mortality rates). Therefore, our findings in Section~\ref{subsec:roc_1} support \cite{Hermann2020} by showing that any policy to control the COVID-19 outbreak successfully has to be an \emph{adaptive} strategy. 
This means we need to constantly measure, monitor and estimate the current numbers and adapt our policy accordingly, i.e., feedback is necessary for reliably handling the outbreak.
\item If we already a priori take into account that our model includes mismatches and that all measured and estimated numbers are not exact and can have a bias, we can further improve the outcome, as shown in Section~\ref{subsec:roc_2}. More specifically, we developed a robust MPC-based feedback strategy using interval arithmetic. The application of feedback without the robust description of the considered model can lead to intermediate increases in the number of new infections necessitating another period of lockdown. On the contrary, a robust feedback strategy can take these model mismatches and other uncertainties into account and is hence able to avoid such behavior, thus significantly reducing the number of fatalities.
\item When looking at the qualitative results the robust MPC-based feedback offers, one can see that, accounting for the instability and uncertainty of the spread of the virus, the controller suggests a rather strict policy at the beginning and only then allows for a gradual increase in the infection rate. Keeping this loosening slow at the beginning shows a beneficial effect in the long run. This qualitative result of the robust MPC underpins also the German policy and reaction to the outbreak of COVID-19 in Germany where initially strong measures (what we here refer to as lockdown) were applied. Only very recently the German government started to loosen these measures slowly and gradually.
\end{itemize}

There are also influences on the course of the outbreak that were not taken into account in the present paper but which are important in an overall strategy towards the spread of COVID-19 (e.g. increasing testing capacities, tracking of infections, as well as investigating which measures lead to the desired infection rate). However, controlling the infection rate is certainly one of the key factors and hence, this paper contributes towards mitigating the spread of COVID-19 under manageable societal and economic costs. We hope that the proposed feedback strategies inspire further investigations in this direction and offer qualitative and high-level insights that underpin the current policies or strategy papers.

\bibliographystyle{elsarticle-num}
\bibliography{Literature}

\begin{thebibliography}{10}
\expandafter\ifx\csname url\endcsname\relax
  \def\url#1{\texttt{#1}}\fi
\expandafter\ifx\csname urlprefix\endcsname\relax\def\urlprefix{URL }\fi
\expandafter\ifx\csname href\endcsname\relax
  \def\href#1#2{#2} \def\path#1{#1}\fi

\bibitem{Maharaj2012}
S.~Maharaj, A.~Kleczkowski, Controlling epidemic spread by social distancing:
  Do it well or not at all, BMC Public Health 12~(679) ({2012}) 1--16.

\bibitem{Kissler2020}
S.~Kissler, C.~Tedijanto, M.~Lipsitch, Y.~H. Grad, Social distancing strategies
  for curbing the {COVID-19} epidemic, medRxiv preprint, (2020).
\newblock \href {http://dx.doi.org/10.1101/2020.03.22.20041079}
  {\path{doi:10.1101/2020.03.22.20041079}}.

\bibitem{Maier2020}
B.~F. Maier, D.~Brockmann, Effective containment explains subexponential growth
  in recent confirmed {COVID-19} cases in {C}hina, Science 368~(6492) (2020)
  742--746.

\bibitem{Dehning2020_change_points_ger}
J.~Dehning, J.~Zierenberg, F.~P. Spitzner, M.~Wibral, J.~P. Neto, M.~Wilczek,
  V.~Priesemann, Inferring covid-19 spreading rates and potential change points
  for case number forecasts, arXiv preprint arXiv:2004.01105, (2020).

\bibitem{Barbarossa2020}
M.~V. Barbarossa, J.~Fuhrmann, J.~Heidecke, H.~V. Varma, N.~Castelletti, J.~H.
  Meinke, S.~Krieg, T.~Lippert, A first study on the impact of current and
  future control measures on the spread of {COVID-19} in {Germany}, medRxiv
  preprint, (2020).
\newblock \href {http://dx.doi.org/10.1101/2020.04.08.20056630}
  {\path{doi:10.1101/2020.04.08.20056630}}.

\bibitem{German2020}
R.~German, A.~Djanatliev, L.~Maile, P.~Bazan, H.~Hackstein, Modeling exit
  strategies from {COVID-19} lockdown with a focus on antibody tests, medRxiv
  preprint, (2020).
\newblock \href {http://dx.doi.org/10.1101/2020.04.14.20063750}
  {\path{doi:10.1101/2020.04.14.20063750}}.

\bibitem{Alleman2020}
T.~Alleman, E.~Torfs, I.~Nopens, {COVID-19}: from model prediction to model
  predictive control, available online
  \url{https://biomath.ugent.be/sites/default/files/2020-04/Alleman_etal_v2.pdf},
  accessed April 30, 2020, (2020).

\bibitem{Nowzari2016}
C.~Nowzari, V.~M. Preciado, G.~J. Pappas, Analysis and control of epidemics,
  IEEE Control Systems Magazine 36 (2016) 26--46.

\bibitem{Casella2020}
F.~{Casella}, Can the {COVID-19} epidemic be controlled on the basis of daily
  test reports?, IEEE Control Systems Letters 5~(3) (2021) 1079--1084.

\bibitem{shorten2020covid}
M.~Bin, P.~Cheung, E.~Crisostomi, P.~Ferraro, C.~Myant, T.~Parisini,
  R.~Shorten, On fast multi-shot epidemic interventions for post lock-down
  mitigation: Implications for simple {COVID}-19 models, arXiv preprint
  arXiv:2003.09930, (2020).

\bibitem{Tsay2020}
C.~Tsay, F.~Lejarza, M.~A. Stadtherr, M.~Baldea, Modeling, state estimation,
  and optimal control for the {US} {COVID-19} outbreak, arXiv preprint
  arXiv:2004.06291, (2020).

\bibitem{Giordano2020}
G.~Giordano, F.~Blanchini, R.~Bruno, P.~Colaneri, A.~D. Filippo, A.~D. Matteo,
  M.~Colaneri, Modelling the {COVID}-19 epidemic and implementation of
  population-wide interventions in {I}taly, Nature Medicine 26 (2020) 855--860.

\bibitem{Demasse2020}
R.~Djidjou-Demasse, Y.~Michalakis, M.~Choisy, M.~T. Sofonea, S.~Alizon, Optimal
  {COVID-19} epidemic control until vaccine deployment, medRxiv preprint,
  (2020).
\newblock \href {http://dx.doi.org/10.1101/2020.04.02.20049189}
  {\path{doi:10.1101/2020.04.02.20049189}}.

\bibitem{kantner2020beyond}
M.~Kantner, T.~Koprucki, Beyond just "flattening the curve": Optimal control of
  epidemics with purely non-pharmaceutical interventions, Journal of
  Mathematics in Industry 10~(1) (2020) 1--23.

\bibitem{Hermann2020}
M.~Meyer-Hermann, I.~Pigeot, V.~Priesemann, A.~Sch\"{o}bel, {Adaptive
  Strategien zur Eind\"{a}mmung der COVID-19-Epidemie},
  {\url{https://www.helmholtz.de/fileadmin/user_upload/01_forschung/28-04-2020_Strategien_zur_Eindaemmung.pdf}},
  accessed April 29, 2020 (2020).

\bibitem{kermack1927contribution}
W.~O. Kermack, A.~G. {McKendrick}, A contribution to the mathematical theory of
  epidemics, Proc. R. Soc. Lond.~(115) (1927) 700--721.

\bibitem{Dong2020_jhu_dashboard}
E.~Dong, H.~Du, L.~Gardner, An interactive web-based dashboard to track
  {COVID}-19 in real time, The Lancet Infectious Diseases 20~(5) (2020)
  533--534.

\bibitem{data_jhu}
E.~Dong, H.~Du, L.~Gardner, 2019 novel coronavirus {COVID}-19 (2019-{nCoV})
  data repository by {J}ohns {H}opkins {CSSE},
  \url{https://github.com/CSSEGISandData/COVID-19}, accessed April 22, 2020
  (2020).

\bibitem{divi2020report0404}
{DIVI-IntensivRegister}, Tagesreport 27.03.2020 -- 21.04.2020, , available
  online: {\url{https://www.divi.de/divi-intensivregister-tagesreport-archiv}}
  (2020).

\bibitem{testPolicyGerZEIT}
J.~W. Jakob~Simmank, Florian~Schumann, So testet {D}eutschland, {ZEIT} online,
  \url{https://www.zeit.de/wissen/gesundheit/2020-03/coronatests-deutschland-coronavirus-covid-19-who-pandemie#wer-soll-in-deutschland-getestet-werden},
  accessed April 15, 2020 (2020).

\bibitem{He2020}
X.~He, E.~H.~Y. Lau, P.~Wu, X.~Deng, J.~Wang, X.~Hao, Y.~C. Lau, J.~Y. Wong,
  Y.~Guan, X.~Tan, X.~Mo, Y.~Chen, B.~Liao, W.~Chen, F.~Hu, Q.~Zhang, M.~Zhong,
  Y.~Wu, L.~Zhao, F.~Zhang, B.~J. Cowling, F.~Li, G.~M. Leung, Temporal
  dynamics in viral shedding and transmissibility of {COVID}-19, Nature
  medicine 26~(5) (2020) 672--675.

\bibitem{Bommer2020}
C.~Bommer, S.~Vollmer, Average detection rate of {SARS-CoV-2} infections is
  estimated around nine percent, available online
  \url{https://www.uni-goettingen.de/en/606540.html}, accessed April 15, 2020
  (2020).

\bibitem{Gudbjartsson2020}
D.~F. Gudbjartsson, A.~Helgason, H.~Jonsson, O.~T. Magnusson, P.~Melsted, G.~L.
  Norddahl, J.~Saemundsdottir, A.~Sigurdsson, P.~Sulem, A.~B. Agustsdottir,
  B.~Eiriksdottir, R.~Fridriksdottir, E.~E. Gardarsdottir, G.~Georgsson, O.~S.
  Gretarsdottir, K.~R. Gudmundsson, T.~R. Gunnarsdottir, A.~Gylfason, H.~Holm,
  B.~O. Jensson, A.~Jonasdottir, F.~Jonsson, K.~S. Josefsdottir,
  T.~Kristjansson, D.~N. Magnusdottir, L.~le~Roux, G.~Sigmundsdottir,
  G.~Sveinbjornsson, K.~E. Sveinsdottir, M.~Sveinsdottir, E.~A. Thorarensen,
  B.~Thorbjornsson, A.~Löve, G.~Masson, I.~Jonsdottir, A.~D. Möller,
  T.~Gudnason, K.~G. Kristinsson, U.~Thorsteinsdottir, K.~Stefansson, Spread of
  {SARS}-{CoV}-2 in the icelandic population, New England Journal of Medicine
  382 (2020) 2302--2315.

\bibitem{Lavezzo2020}
E.~Lavezzo, E.~Franchin, C.~Ciavarella, G.~Cuomo-Dannenburg, L.~Barzon,
  C.~Del~Vecchio, L.~Rossi, R.~Manganelli, A.~Loregian, N.~Navarin, et~al.,
  Suppression of a {SARS}-{CoV}-2 outbreak in the italian municipality of
  vo’, Nature 584~(7821) (2020) 425--429.

\bibitem{Mizumoto2020}
K.~Mizumoto, K.~Kagaya, A.~Zarebski, G.~Chowell, Estimating the asymptomatic
  proportion of coronavirus disease 2019 ({COVID}-19) cases on board the
  {D}iamond {P}rincess cruise ship, {Y}okohama, {J}apan, Eurosurveillance
  25~(10) (2020) 2000180.

\bibitem{Heiden2020}
M.~an~der Heiden, O.~Hamouda, Sch{\"a}tzung der aktuellen {E}ntwicklung der
  {SARS-CoV-2}-epidemie in {D}eutschland - {N}owcasting, Epidemiologisches
  Bulletin 2020~(17) (2020) 10--15.

\bibitem{WHO_report_2020}
{World Health Organization (WHO)}, Report of the {WHO-C}hina joint mission on
  coronavirus disease 2019 ({COVID-19}), available online
  \url{https://www.who.int/docs/default-source/coronaviruse/who-china-joint-mission-on-covid-19-final-report.pdf},
  accessed April 15, 2020, (2020).

\bibitem{Linton2020}
N.~M. Linton, T.~Kobayashi, Y.~Yang, K.~Hayashi, A.~R. Akhmetzhanov, S.~mok
  Jung, B.~Yuan, R.~Kinoshita, H.~Nishiura, Incubation period and other
  epidemiological characteristics of 2019 novel coronavirus infections with
  right truncation: A statistical analysis of publicly available case data,
  Journal of Clinical Medicine 9~(2) (2020) 538.

\bibitem{Li2020}
Q.~Li, X.~Guan, P.~Wu, X.~Wang, L.~Zhou, Y.~Tong, R.~Ren, K.~S. Leung, E.~H.
  Lau, J.~Y. Wong, X.~Xing, N.~Xiang, Y.~Wu, C.~Li, Q.~Chen, D.~Li, T.~Liu,
  J.~Zhao, M.~Liu, W.~Tu, C.~Chen, L.~Jin, R.~Yang, Q.~Wang, S.~Zhou, R.~Wang,
  H.~Liu, Y.~Luo, Y.~Liu, G.~Shao, H.~Li, Z.~Tao, Y.~Yang, Z.~Deng, B.~Liu,
  Z.~Ma, Y.~Zhang, G.~Shi, T.~T. Lam, J.~T. Wu, G.~F. Gao, B.~J. Cowling,
  B.~Yang, G.~M. Leung, Z.~Feng, Early transmission dynamics in {W}uhan,
  {C}hina, of novel coronavirus{\textendash}infected pneumonia, New England
  Journal of Medicine 382~(13) (2020) 1199--1207.

\bibitem{Yang2020}
X.~Yang, Y.~Yu, J.~Xu, H.~Shu, J.~Xia, H.~Liu, Y.~Wu, L.~Zhang, Z.~Yu, M.~Fang,
  T.~Yu, Y.~Wang, S.~Pan, X.~Zou, S.~Yuan, Y.~Shang, Clinical course and
  outcomes of critically ill patients with {SARS}-{CoV}-2 pneumonia in {W}uhan,
  {C}hina: a single-centered, retrospective, observational study, The Lancet
  Respiratory Medicine 8 (2020) 475--481.

\bibitem{Wang2020}
D.~Wang, B.~Hu, C.~Hu, F.~Zhu, X.~Liu, J.~Zhang, B.~Wang, H.~Xiang, Z.~Cheng,
  Y.~Xiong, Y.~Zhao, Y.~Li, X.~Wang, Z.~Peng, Clinical characteristics of 138
  hospitalized patients with 2019 novel coronavirus{\textendash}infected
  pneumonia in {W}uhan, {C}hina, {JAMA} 323~(11) (2020) 1061--1069.

\bibitem{andersson2019casadi}
J.~A. Andersson, J.~Gillis, G.~Horn, J.~B. Rawlings, M.~Diehl, {CasADi}: a
  software framework for nonlinear optimization and optimal control,
  Mathematical Programming Computation 11~(1) (2019) 1--36.

\bibitem{verity2020estimates}
{R. Verity et al.}, Estimates of the severity of coronavirus disease 2019: a
  model-based analysis, The Lancet Infectious Diseases 20~(6) (2020) 669 --
  677.

\bibitem{grasselli2020critical}
G.~Grasselli, A.~Pesenti, M.~Cecconi, Critical care utilization for the
  {COVID-19} outbreak in {Lombardy}, {Italy}: early experience and forecast
  during an emergency response, {JAMA} 323~(16) (2020) 1545--1546.

\bibitem{IMHE2020forecasting}
{IHME COVID-19 health service utilization forecasting team, C. J. L. Murray},
  Forecasting {COVID-19} impact on hospital bed-days, {ICU}-days,
  ventilator-days and deaths by {US} state in the next 4 months, medRxiv
  preprint, (2020).
\newblock \href {http://dx.doi.org/10.1101/2020.03.27.20043752}
  {\path{doi:10.1101/2020.03.27.20043752}}.

\bibitem{ji2020potential}
Y.~Ji, Z.~Ma, M.~P. Peppelenbosch, Q.~Pan, Potential association between
  {COVID-19} mortality and health-care resource availability, The Lancet Global
  Health 8~(4) (2020) e480.

\bibitem{iss2020report}
{COVID-19 Surveillance Group}, Characteristics of {COVID-19} patients dying in
  {Italy}, report based on available data on {March} 30th, 2020, available
  online:
  {\url{https://www.epicentro.iss.it/coronavirus/bollettino/Report-COVID-2019_30_marzo_eng.pdf}},
  accessed April 16, 2020, (2020).

\bibitem{RKI2017}
{Robert Koch Institut}, {E}pidemiologisches {B}ulletin {N}r.27, p.~243,
  available online
  \url{https://www.rki.de/DE/Content/Infekt/EpidBull/Archiv/2017/Ausgaben/27_17.pdf?__blob=publicationFile}
  (2017).

\bibitem{Helmholtz2020}
{Helmholtz-Initiative 'Systemische Epidemiologische Analyse der
  Covid-19-Epidemie'}, {S}tellungnahme der {H}elmholtz-{I}nitiative
  '{Systemische Epidemiologische Analyse der COVID-19-Epidemie}',
  {\url{https://www.helmholtz.de/fileadmin/user_upload/01_forschung/Helmholtz-COVID-19-Papier_02.pdf}},
  accessed April 15, 2020, (2020).

\bibitem{Leopoldina2020}
Leopoldina, Dritte {A}d-hoc-{S}tellungnahme: Coronavirus-{P}andemie - {D}ie
  {K}rise nachhaltig \"{u}berwinden, available online
  {\url{https://www.leopoldina.org/uploads/tx_leopublication/2020_04_13_Coronavirus-Pandemie-Die_Krise_nachhaltig_%C3%BCberwinden_final.pdf}},
  accessed April 15, 2020, (2020).

\bibitem{preciado2014optimal}
V.~M. Preciado, M.~Zargham, C.~Enyioha, A.~Jadbabaie, G.~J. Pappas, Optimal
  resource allocation for network protection against spreading processes, IEEE
  Transactions on Control of Network Systems 1~(1) (2014) 99--108.

\bibitem{kohler2018dynamic}
J.~K{\"o}hler, C.~Enyioha, F.~Allg{\"o}wer, Dynamic resource allocation to
  control epidemic outbreaks a model predictive control approach, in: Proc.
  Annual American Control Conference (ACC), 2018, pp. 1546--1551.

\bibitem{Dorfman1958}
R.~Dorfman, P.~A. Samuelson, R.~M. Solow, Linear programming and economic
  analysis, Courier Corporation, 1987.

\bibitem{Oliver2020}
N.~Oliver, E.~Letouz\'e, H.~Sterly, S.~Delataille, M.~D. Nadai, B.~Lepri,
  R.~Lambiotte, R.~Benjamins, C.~Cattuto, V.~Colizza, N.~de~Cordes, S.~P.
  Fraiberger, T.~Koebe, S.~Lehmann, J.~Murillo, A.~Pentland, P.~N. Pham,
  F.~Pivetta, A.~A. Salah, J.~Saram\"aki, S.~V. Scarpino, M.~Tizzoni,
  S.~Verhulst, P.~Vinck, Mobile phone data and {COVID}-19: Missing an
  opportunity?, (2020).
\newblock \href {http://arxiv.org/abs/2003.12347v1}
  {\path{arXiv:2003.12347v1}}.

\bibitem{rawlings2017model}
J.~B. Rawlings, D.~Q. Mayne, M.~Diehl, Model Predictive Control: Theory,
  Computation, and Design, Nob Hill Publishing, 2017.

\bibitem{limon2005robust}
D.~Limon, J.~Bravo, T.~Alamo, E.~Camacho, Robust {MPC} of constrained nonlinear
  systems based on interval arithmetic, IEE Proceedings-Control Theory and
  Applications 152~(3) (2005) 325--332.

\bibitem{watkins2019robust}
N.~J. Watkins, C.~Nowzari, G.~J. Pappas, Robust economic model predictive
  control of continuous-time epidemic processes, IEEE Transactions on Automatic
  Control 65 (2019) 1116--1131.

\end{thebibliography}
\appendix
\section{Optimal control formulation using terminal constraints}
\label{app:OC_term}
In order to avoid artifacts of considering a finite-horizon problem (e.g. a lot of infected people at the end of the horizon), an alternative to considering the modified cost function F from~\eqref{eq:terminal_cost_function} is the inclusion of additional terminal constraints for the contagious population $IDART=(I,D,A,R,T)\in\mathbb{R}^{5}$.
In particular, we require that at the end of the control horizon $N$, the number of contagious individuals in each category ($I,D,A,R,T$) should  be smaller than the corresponding number from the baseline policy (c.f.~\eqref{eq:OC_term_1}). 
In addition,  at the end of the horizon the number of contagious individuals should be non-increasing, which is implemented as~\eqref{eq:OC_term_2}.  

\begin{figure}
\begin{center}
\includegraphics[width=0.75\textwidth]{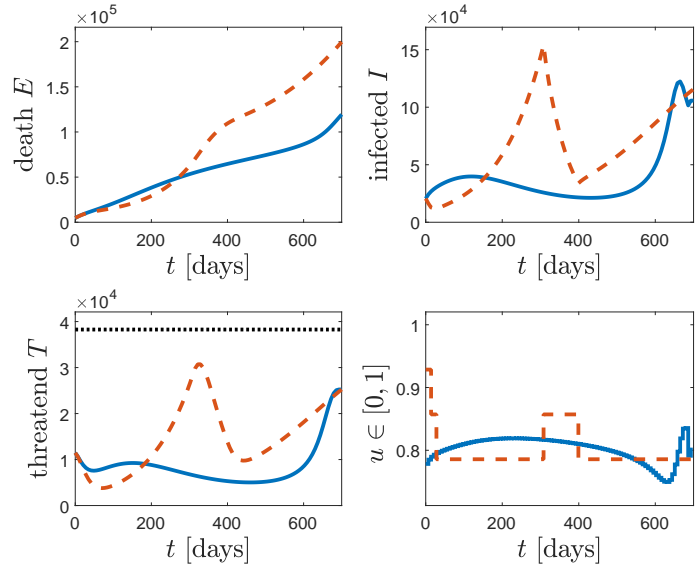} 
\caption{Cautious baseline: Optimal control strategy with terminal constraints (blue, solid), baseline  policy (red, dashed) and ICU capacity (dotted, black).}  \label{fig:OC_term_1}
\end{center}
\end{figure}

\begin{figure}
\begin{center}
\includegraphics[width=0.75\textwidth]{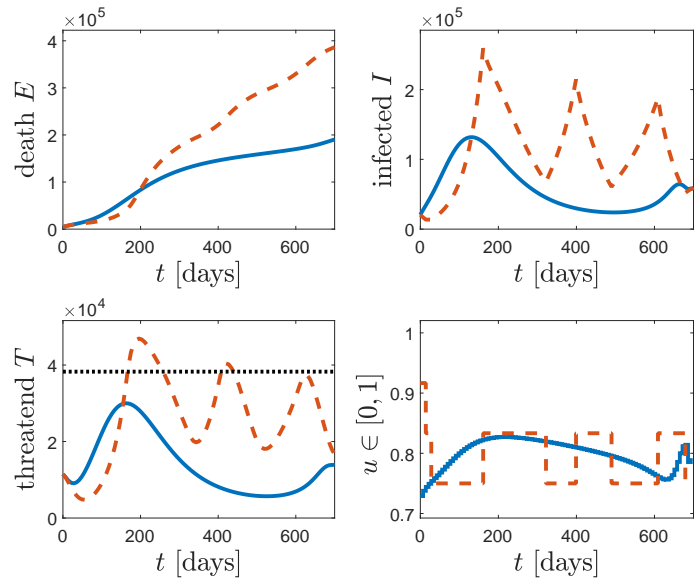} 
\caption{Aggressive baseline: Optimal control strategy with terminal constraints (blue, solid), baseline policy (red, dashed) and ICU capacity (dotted, black).} \label{fig:OC_term_2}
\end{center}
\end{figure}

Hence, in the following, we replace the cost $F$ by the number of fatalities $E$, and we add the following constraints to the optimal control problem \eqref{eq:OC} from Section~\ref{sec:OC_3}:
\begin{subequations}
	\begin{align}
	\label{eq:OC_term_1}
	&IDART(N\cdot T_s)\leq IDART^b(N\cdot T_s),\\
	\label{eq:OC_term_2}
	&IDART(N\cdot T_s)\leq IDART((N-1)\cdot T_s).
	\end{align}
\end{subequations}
Again, the index $k$ in~\eqref{eq:OC} corresponds to weeks and the states $IDART(k\cdot T_s)$ correspond to the result of simulating the system~\eqref{eq:model} with the parameters and initial condition from Section~\ref{sec:modeling}. 
These terminal conditions~\eqref{eq:OC_term_1}--\eqref{eq:OC_term_2} (which should be interpreted element-wise) ensure that the final state after the finite horizon $N$ is ``better'' than the baseline solution (c.f.~\eqref{eq:OC_term_1}) and the outbreak can be contained (c.f.~\eqref{eq:OC_term_2}).
The simulation results with the two baseline policies shown in Figure \ref{fig:OC_term_1} and \ref{fig:OC_term_2} demonstrate that the terminal constraints indeed effectively prevent the turnpike behavior.
However, the additional constraints also lead to a slight increase in the number of fatalities.


\section{Alternative average constraint formulation}
\label{app:OC} 
\begin{figure}
\begin{center}
\includegraphics[width=0.75\textwidth]{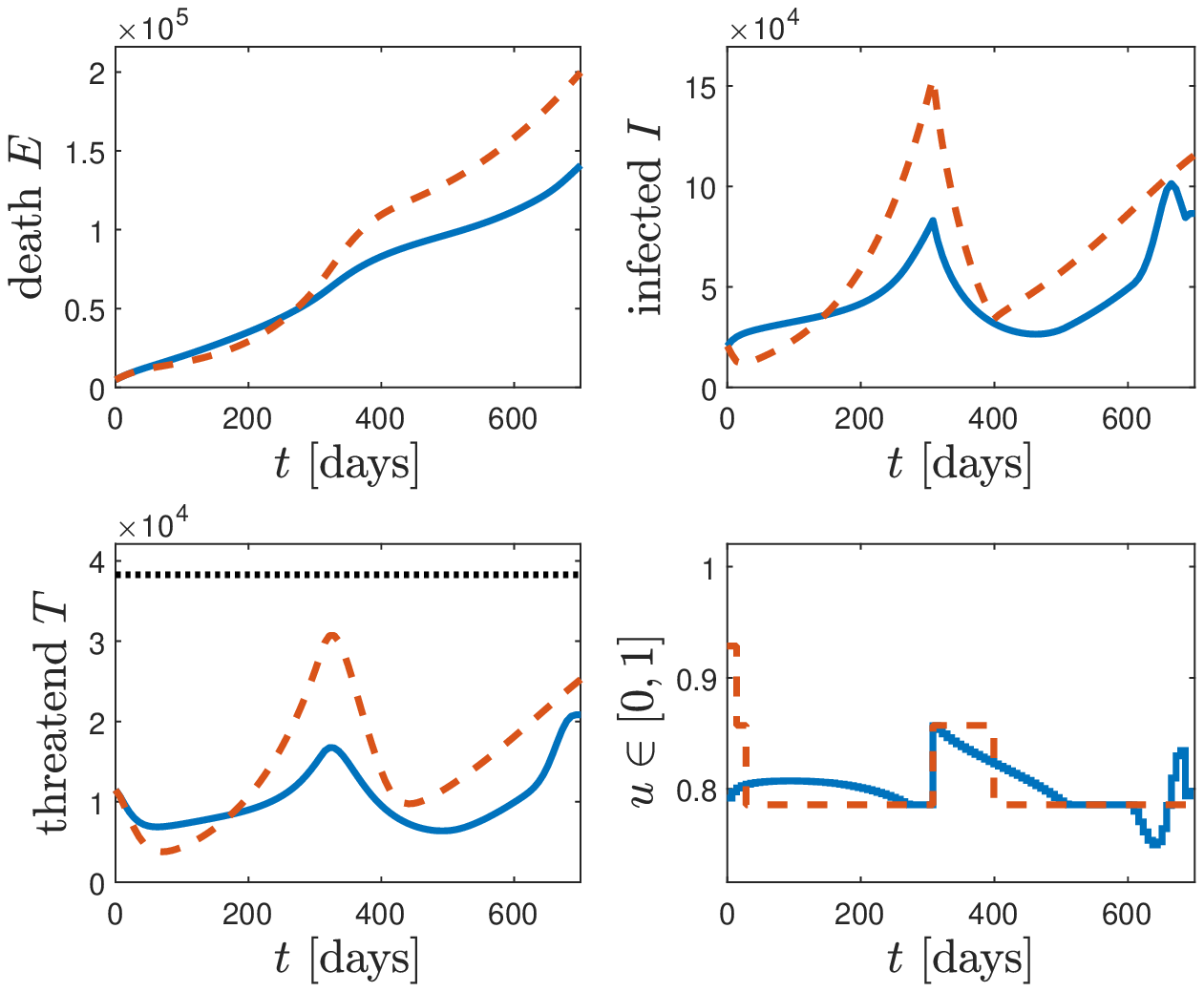} 
\includegraphics[width=0.75\textwidth]{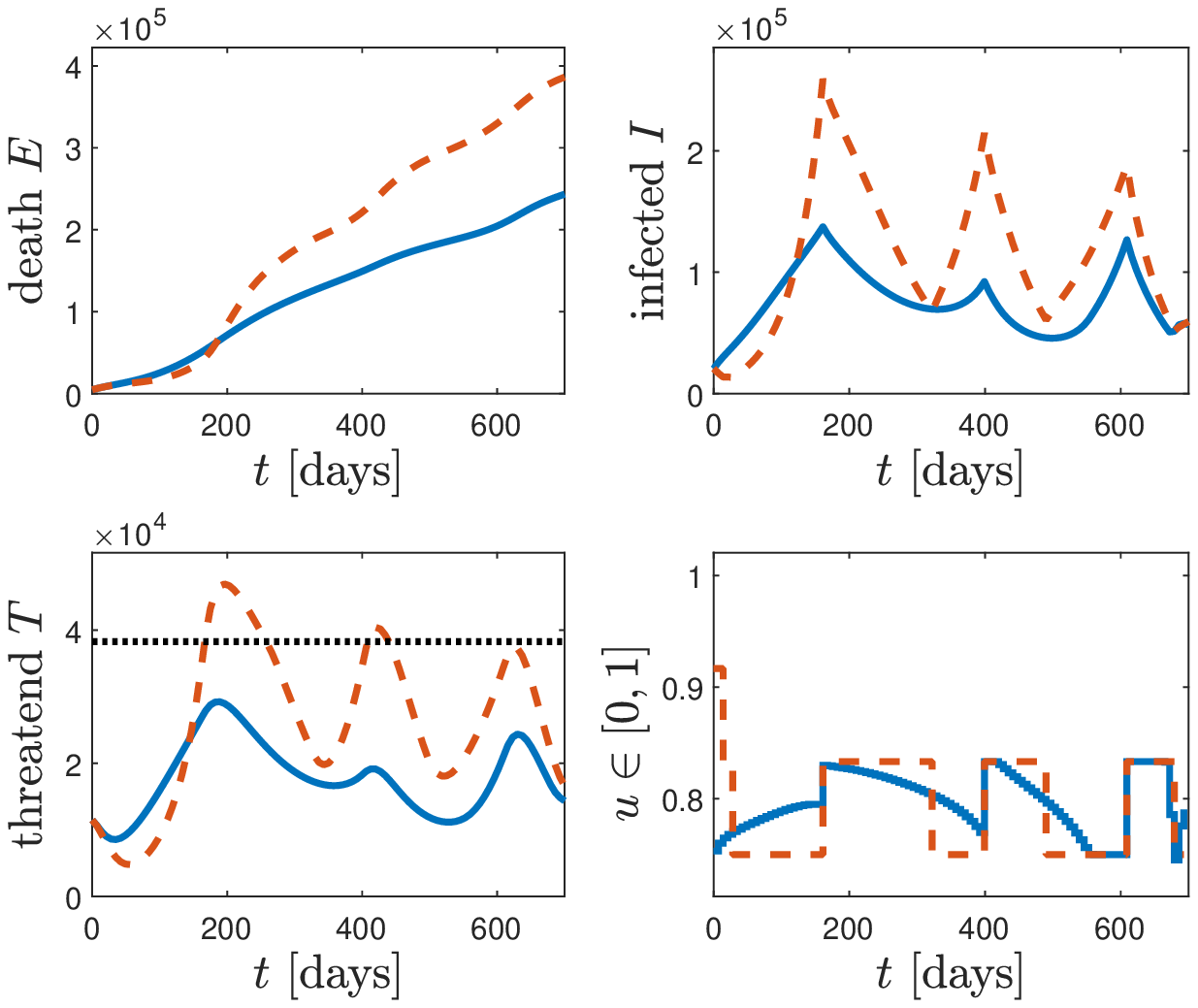} 
\end{center}
\caption{Transient constraint on social policy: Top: cautious baseline; Bottom: aggressive baseline.
Optimal control strategy (blue, solid), baseline policy (red, dashed) and ICU capacity (dotted, black).}
\label{fig:OC_app}
\end{figure}

We wish to briefly mention a stronger restriction on the societal  cost of the optimal control strategy. 
In particular, instead of only restricting the cost over the considered horizon of $N=100$ weeks, a stronger property is to ensure that at any time $t$, the previously accumulated policy cost is smaller than the corresponding cost of the baseline policy. 
This can be done by replacing condition~\eqref{eq:OC_policy} with the following transient constraint:
\begin{align}
&\sum_{k=0}^{i-1}c_{\text{policy}}(u_1(k))\leq \sum_{k=0}^{i-1}c_{\text{policy}}(u^b_1(k)),\\
&i=1,\dots,N.
\end{align}
The corresponding results for both baselines considered in Section~\ref{sec:OC_3} can be seen in Figure~\ref{fig:OC_app}.
In this case the number of fatalities are reduced by $33\%$ and $37\%$, respectively.
Thus, also for this more restrictive setting, the optimal controller can significantly reduce the number of fatalities. 
In addition, in the comparison to the more aggressive baseline we also see that \textit{early} measures are absolutely crucial, since the two policies are essentially equivalent in the period from 110 days--150 days, i.e., including the critical time period where the ICU capacity is exceeded, but differ significantly over 30 weeks prior to the violation of the ICU capacity.

\section{Interval predictions}
\label{app:interval}
In the following, we derive the dynamics of the interval predictions $\overline{f},\underline{f}$ used in~\eqref{eq:interval_predict}. 
 Note that the following property holds for any scalars $a\in[\underline{a},\overline{a}]$, $b\in[\underline{b},\overline{b}]$:
 \begin{align}
a\cdot b\in[\underline{a}\underline{b},\overline{a}\overline{b}],
\label{eq:condition_1}
\end{align}
if $\underline{a},\underline{b}\geq 0$. 
Furthermore, to ensure that $x(t)\in[\underline{x}(t),\overline{x}(t)]$ given $x(0)\in[\underline{x}(0),\overline{x}(0)]$, we only require 
\begin{align}
\dot{\overline{x}}_i\geq \dot{x}_i \quad \text{if} \>\> \overline{x}_i=x_i,
\label{eq:condition_2}
\end{align}
 and similarly 
\begin{align}
\dot{\underline{x}}_i\leq \dot{x}_i \quad \text{if} \>\> \underline{x}_i=x_i,
\label{eq:condition_3}
\end{align}
for all $i=1, \dots, 8$.
Essentially, we use the property \eqref{eq:condition_1} together with the fact that $x_i$ and the parameters are positive to ensure that \eqref{eq:condition_2}and \eqref{eq:condition_3} hold in order to derive the differential equations for the interval. More precisely, assuming that every parameter is uncertain in some bounds (e.g. $\beta\in[\underline{\beta},\overline{\beta}]$) yields the following $2\cdot 8$ ODEs:
\begin{subequations}\label{eq:model_interval}
\begin{align}
\label{eq:model_S_interval}
\dot{\overline{S}}=&-\overline{S}(\underline{\alpha}\underline{I}+\underline{\beta}\underline{D}+\underline{\gamma}\underline{A}+\underline{\beta}\underline{R}),\\
\label{eq:model_S_interval2}
\dot{\underline{S}}=&-\underline{S}(\overline{\alpha}\overline{I}+\overline{\beta}\overline{D}+\overline{\gamma}\overline{A}+\overline{\beta}\overline{R}),\\
\dot{\overline{I}} =&\overline{S}(\overline{\alpha}\overline{I}+\overline{\beta}\overline{D}+\overline{\gamma}\overline{A}+\overline{\beta}\overline{R})-(\underline{\epsilon}+\underline{\zeta}+\underline{\lambda})\overline{I},\\
\dot{\underline{I}} =&\underline{S}(\underline{\alpha}\underline{I}+\underline{\beta}\underline{D}+\underline{\gamma}\underline{A}+\underline{\beta}\underline{R})-(\overline{\epsilon}+\overline{\zeta}+\overline{\lambda})\underline{I},\\
\dot{\overline{D}}=&\overline{\epsilon}\overline{I}-(\underline{\zeta}+\underline{\lambda})\overline{D},\\
\dot{\underline{D}}=&\underline{\epsilon}\underline{I}-(\overline{\zeta}+\overline{\lambda})\underline{D},\\
\dot{\overline{A}}=&\overline{\zeta}\overline{I}-(\underline{\theta}+\underline{\mu}+\underline{\kappa})\overline{A},\\
\dot{\underline{A}}=&\underline{\zeta}\underline{I}-(\overline{\theta}+\overline{\mu}+\overline{\kappa})\underline{A},\\
\dot{\overline{R}}=&\overline{\zeta}\overline{D}+\overline{\theta}\overline{A}-(\underline{\mu}+\underline{\kappa})\overline{R},\\
\dot{\underline{R}}=&\underline{\zeta}\underline{D}+\underline{\theta}\underline{A}-(\overline{\mu}+\overline{\kappa})\underline{R},\\
\dot{\overline{T}}=&\overline{\mu}\overline{A}+\overline{\mu}\overline{R}-(\underline{\sigma}(\overline{T})+\underline{\tau}(\overline{T}))\overline{T},\\
\dot{\underline{T}}=&\underline{\mu}\underline{A}+\underline{\mu}\underline{R}-(\overline{\sigma}(\underline{T})+\overline{\tau}(\underline{T}))\underline{T},\\
\dot{\overline{H}}=&\overline{\lambda}\overline{I}+\overline{\lambda}\overline{D}+\overline{\kappa}\overline{A}+\overline{\kappa}\overline{R}+\overline{\sigma}(\overline{T})\overline{T},\\
\dot{\underline{H}}=&\underline{\lambda}\underline{I}+\underline{\lambda}\underline{D}+\underline{\kappa}\underline{A}+\underline{\kappa}\underline{R}+\underline{\sigma}(\underline{T})\underline{T},\\
\dot{\overline{E}}=&\overline{\tau}(\overline{T})\overline{T},\\
\dot{\underline{E}}=&\underline{\tau}(\underline{T})\underline{T}.
\end{align}
\end{subequations}
Since these dynamics only correspond to possibly conservative overapproximations, we can use $\sum_{i=1}^8 x_i=1$  to possibly improve the resulting bounds for $S$ using the following projections:
$\overline{S}\leq 1-\sum_{i=2}^8\underline{x}_i$ and $\underline{S}\geq 1-\sum_{i=2}^8\overline{x}_i$. 
In principle it would also be possible to directly set $\overline{S},\underline{S}$ using the other states $\overline{x}_i,\underline{x}_i$ instead of simulating~\eqref{eq:model_S_interval}--\eqref{eq:model_S_interval2}, but this may not necessarily ensure $\overline{S}\leq 1$ and $\underline{S}\geq 0$.

\end{document}